\documentclass[11pt]{article}
\usepackage{sectsty}
\usepackage{graphicx}
\usepackage[utf8]{inputenc}
\usepackage{graphicx}
\usepackage[utf8]{inputenc}
\usepackage{svg}
\usepackage{hyperref}
\usepackage{float}
\usepackage{amsmath}
\usepackage{amsthm}
\usepackage{amsthm}
\usepackage{natbib}
\usepackage{lipsum}
\usepackage{setspace}
\usepackage{nccmath}
\usepackage{changes}
\usepackage{mathtools}
\usepackage{enumerate}
\usepackage{amssymb}
\usepackage{tabularx}
\usepackage{rotating}
\newtheorem{definition}{Definition}
\title{A Complete Characterisation of Structured Missingness}
\author{James Jackson$^{1}$, Robin Mitra$^{1,2}$, Niels Hagenbuch$^{3}$, Sarah McGough$^{4}$, Chris Harbron$^{5}$}
\newcommand\independent{\protect\mathpalette{\protect\independenT}{\perp}}
\def\independenT#1#2{\mathrel{\rlap{$#1#2$}\mkern2mu{#1#2}}}
\newcommand{\pkg}[1]{{\normalfont\fontseries{b}\selectfont #1}}
\newcommand{\noop}[1]{}

\begin{document} 
\maketitle	
\begin{center}
    {$^1$The Alan Turing Institute, London, UK, $^2$Department of Statistical Science, University College London, London, UK, $^3$F. Hoffmann-La Roche AG, Basel, Switzerland, $^4$Genentech, South San Francisco, CA, USA, $^5$Roche Pharmaceuticals, Welwyn Garden City, UK}
\end{center}
\begin{center}
    {jjackson@turing.ac.uk}
\end{center}
\begin{abstract}
{Our capacity to process large complex data sources is ever-increasing, providing us with new, important applied research questions to address, such as how to handle missing values in large-scale databases. \cite{Mitra2023} noted the phenomenon of Structured Missingness (SM), which is where missingness has an underlying structure. Existing taxonomies for defining missingness mechanisms typically assume that variables' missingness indicator vectors $M_1,\;M_2, \hdots, M_p$ are independent after conditioning on the relevant portion of the data matrix $\mathbf{X}$. As this is often unsuitable for characterising SM in multivariate settings, we introduce a taxonomy for SM, where each ${M}_j$ can depend on $\mathbf{M}_{-j}$ (i.e., all missingness indicator vectors except ${M}_j$), in addition to $\mathbf{X}$. We embed this new framework within the well-established decomposition of mechanisms into MCAR, MAR, and MNAR \citep{Rubin1976}, allowing us to recast mechanisms into a broader setting, where we can consider the combined effect of $\mathbf{X}$ and $\mathbf{M}_{-j}$ on ${M}_j$. We also demonstrate, via simulations, the impact of SM on inference and prediction, and consider contextual instances of SM arising in a de-identified nationwide (US-based) clinico-genomic database (CGDB). We hope to stimulate interest in SM, and encourage timely research into this phenomenon.}
\end{abstract}
\section{Introduction}
    {Missing values are to be expected in most data sets, especially when the underlying data collection process is complex, such as in a clinical setting. The underlying mechanism driving the missingness dictates whether adjustments are required in order to obtain valid inferences. Over the years a substantial amount of literature has been devoted to dealing with missing data \citep{Rubin1987, Enders2010, OKelly2014, Molenberghs2014, Little2019, Carpenter2023}.} \par The {seminal} taxonomy of missing completely at random (MCAR), missing at random (MAR), and missing not at random (MNAR), introduced by \cite{Rubin1976}, {}{has been fundamental to {modelling} missingness mechanisms}. It allow{s} analysts to understand the relationship between missingness and the observed and missing portions of the data matrix $\mathbf{X}$, {}{thus allowing the missing data to be appropriately addressed.} \par It is unclear, however, how this taxonomy deals with the case where there is an underlying structure to the missingness itself. Firstly, it assumes an element of randomness {and does not cover the scenario where values are missing with certainty. Secondly, after conditioning upon relevant variables, \citeauthor{Rubin1976}'s taxonomy assumes the probability of missingness is equal across all subjects. That is, i}n the MCAR case, it is assumed that, for a given variable, missing values are equally likely across all subjects; similarly, in the MAR case, it is assumed {that }missing values are equally likely across all subjects after conditioning on the relevant portion of the observed data; and a similar notion holds in the case of MNAR. Moreover, MCAR, MAR, and MNAR assume {that} the {rows and columns of the }missingness indicator matrix -- a matrix comprising of zeros and ones denoting whether each value in the data matrix is observed or missing -- are conditionally independent. Yet, such assumptions do not always hold. What if missingness is non-random? What if dependencies exist among the missingness indicators themselves? What if factors other than the data are driving the missingness? This is where the notion of structured missingness (SM) comes in. \par 
The challenges associated with SM were recently set out by \cite{Mitra2023}. To summarise briefly, SM is an umbrella term covering a range of missingness mechanisms that have an underlying structure. \citeauthor{Mitra2023} list five ``common routes to SM'', demonstrating how SM can arise in practice. The first two of these relate to data linkage -- specifically, to multi-modal and multi-scale linkage -- which often leads to large swathes of missing data that is not random but rather a certainty. The third route to SM is through batch failure, which can result in missingness with a sequential aspect, and the last two relate to skip patterns and population heterogeneity.  \par Research into SM is important for a number of reasons. {Strong dependencies between variables' missingness indicator vectors can cause inferential challenges; for example, it may not be possible to fit imputation models to the data. Another reason is that missingness structures themselves can potentially hold useful information; for example, an interaction detected in the missingness indicators for two variables, may reveal an insight into the relationship between those variables, a relationship that may not be obvious from the data values.} In this respect, exploring SM shares similarities with the use of paradata \citep{Couper1998}, to fully extract all available information from a data set. 

{Problems due to SM have been previously encountered when dealing with incomplete data. 
However, the literature tends to be scattered, focusing on specific instances of SM without a wider appreciation of the phenomenon, as well as tending to re-purpose existing methods rather than developing bespoke framework that SM clearly merits. 
\cite{tierney2015} propose utilising tree models to learn about structures present among the missing data but do not elaborate further on this idea. \cite{mohan2021} consider graphical models to represent multivariate dependencies in incomplete data, with the potential to incorporate relationships between missingness indicators, although their focus is primarily on addressing causal inference problems and not SM. Modelling and imputation methods that handle, among other things, certain types of SM have also been developed; for example, \cite{audigier2018} developed a multi-level model for incomplete data, and \cite{Dong2019} proposed a non-parametric imputation model for blocks of missing values. While these approaches may be adequate for the specific problems they seek to address, they do not provide a complete picture of the rich landscape SM describes, and consequently may belie some of the deeper questions it poses.}  

\par   For further work to proceed in the area of SM, we need to {first }define what a SM mechanism means. {{\cite{Mitra2023} introduced nine Grand Challenges to be addressed in relation to SM, of which defining SM is the first. It is also arguably the most pressing, as it is intrinsic to the other grand challenges and the direction of future research.}} {In this paper, we identify and define SM mechanisms relating to multivariate missingness structures; that is, we focus on {}{relationships} among missingness indicator variables themselves, {}{as well as} relationships between missingness indicators and variables in the data. {By doing so, we introduce a taxonomy that comprehensively details the mathematical framework underpinning SM, and embed this within \citeauthor{Rubin1976}'s taxonomy, thus generalising the concept of a missing data mechanism to better deal with complex multivariate incomplete data settings. {}{The purpose of this paper is not, therefore, to offer solutions for SM, but rather to provide a solid foundation that defines and characterises this phenomenon, by considering the myriad ways it can manifest.}}} \par
{}{The remainder of this paper is organised as follows. Section \ref{sec2} revisits \citeauthor{Rubin1976}'s taxonomy, and Section \ref{sec3} explores its limitation in relation to SM. {In} Section \ref{sec4}, {we} detail our SM taxonomy, structuring this to be aligned with the classic MCAR/MAR/MNAR decomposition of missing data mechanisms. Section \ref{sec7} covers remaining forms of SM that fall outside the categorisation given in Section \ref{sec4}.} {}{Section \ref{sec8} provides {motivating simulation examples that illustrate the effects of SM on analyses and inferential validity.} Section \ref{sec9} {}{provides contextual settings where SM mechanisms can arise, using a large Clinico-Genomic Database (CGDB) for illustration}. Section \ref{sec10} {}{ends with some} concluding remarks.

\section{\citeauthor{Rubin1976}'s taxonomy} \label{sec2}

{}{To formally define SM, we return to the}
definitions introduced by \cite{Rubin1976}, widely known as MCAR, MAR, and MNAR. To define these, suppose we have an {{individual-level }}data set $\mathbf{X} = (X_1,X_2,\hdots,X_p)$ comprising $n$ subjects ($n$ rows) and $p$ variables ($p$ columns), such that the data form an $n\times p$ matrix. Suppose we have a corresponding $n\times p$ missingness indicator matrix $\mathbf{M} = (M_1,M_2,\hdots,M_p)$, where ${M_{ij}} = 1$ if observation $X_{ij}$ is missing and $M_{ij} = 0$ if $X_{ij}$ is observed. We can then decompose $\mathbf{X}$ into its observed and missing portions, $\mathbf{X}_\text{obs}$ and $\mathbf{X}_\text{mis}$, respectively:

\begin{align}
\mathbf{X}_\text{obs} =\{X_{ij} \mid M_{ij} =0\} \quad \text{and} \quad \mathbf{X}_\text{mis} =\{X_{ij} \mid M_{ij} =1\}. \nonumber
\end{align}

It is typical in the missing data literature to conceptualise the missingness mechanism as a binary probability function $p(\mathbf{M}\mid \mathbf{X}, \gamma)$, where $\gamma$ is a parameter space governing $\mathbf{M}$. The matrix $\mathbf{M}$ has the same dimensions as the data matrix $\mathbf{X}$, so whenever $\mathbf{X}$ is multivariate $(p>1)$, $\mathbf{M}$ is also multivariate. \par
Importantly, \citeauthor{Rubin1976}'s definitions of MCAR, MAR, and MNAR {}{do not consider the relationship between $M_j$, the missingness indicator vector for the $j$th variable, and $\mathbf{M}_{-j}$, the missingness indicator vectors for all variables except the $j$th. In effect, it is implicitly assumed that $M_j$ and $\mathbf{M}_{-j}$ are {(conditionally)} independent (for all $j\in \{1, \hdots, p \}$). This independence can be explicitly shown in the definitions, by expressing $M_j$ in terms of} $\mathbf{X}$, $\gamma$, {and now {\em additionally} $\mathbf{M}_{-j}$}. 
\begin{definition} \normalfont\label{def2.1}
Data in $\mathbf{X}$ are missing completely at random (MCAR) if{}{, for each variable $M_j$},
\begin{align*}
    p({M}_j\mid \mathbf{M}_{-j}, \mathbf{X}, \gamma) &= p({M}_j\mid\gamma) \quad \forall \; j, \mathbf{X}, \gamma.
\end{align*}
\end{definition}
\begin{definition} \normalfont\label{def2.2}
Data in $\mathbf{X}$ are missing at random (MAR) if{}{, for each variable $M_j$},
\begin{align*} 
    p({M}_j\mid \mathbf{M}_{-j}, \mathbf{X}, \gamma) &= p({M}_j\mid\mathbf{X}_\text{obs}, \gamma) \quad \forall\; j,\mathbf{X}_\text{mis}, \gamma.
\end{align*}
\end{definition}
\begin{definition} \normalfont\label{def2.3}
Data in $\mathbf{X}$ are missing not at random (MNAR) if{}{, for each variable $M_j$},
\begin{align*}
    p({M}_j\mid \mathbf{M}_{-j}, \mathbf{X}, \gamma) &= p({M}_j\mid\mathbf{X}_\text{obs}, \mathbf{X}_\text{mis}, \gamma) \quad \forall\;j, \gamma. 
\end{align*}
\end{definition}
{Here, $\mathbf{X}_\text{mis}$ also includes variables not necessarily included in the data set, such as an unobserved or latent variable.} {While traditionally $\mathbf{M}_{-j}$ is not explicitly conditioned on in the mechanism, we can see from the above definitions that the different mechanisms correspond to {independence assumptions between ${M}_{j}$ and $\mathbf{M}_{-j}$, described in more detail in Section 3.2.} 
\par When modelling the joint distribution of {a} multivariate data set, it is common to decompose it into a product of conditional univariate distributions, as is done, for example, to $\mathbf{X}$ when fitting imputation models. Thus, we can express $\mathbf{M}$ in the following form:    
\begin{align}
    p(\mathbf{M}) &=
p(M_1, M_2, \hdots, M_p) = p(M_1) \prod_{j=2}^p p(M_{j} \mid M_{j-1}, \hdots, M_1). \label{eq1}
\end{align} Given this formulation, it would suffice to say that ${M}_j$ is independent of $M_1, \hdots, M_{j-1}$ (rather than $\mathbf{M}_{-j}$) for all $j$. \par The next section delves into the limitations of this taxonomy, thus paving the way for a new taxonomy relating to SM.
\section{Limitations with existing terminology} \label{sec3}
Here, we describe two primary limitations of existing terminology in relation to SM, thereby motivating the need for further terminology.
\subsection{Non-random missingness}
The definitions of MCAR, MAR, and MNAR assume that missingness, to a certain extent, is random. In the traditional domain of survey data sets, where subjects are selected at random and are then either missing or observed at random, this notion of random missingness suffices. Yet in many complex data sets -- where individuals are neither selected at random nor are missing at random -- the definition begins to break down. Surveys, which are designed with statistical analysis in mind, typically ensure that subjects are randomly selected from a well-defined sampling frame that is representative of the population at large. In a non-survey environment, however, data do not necessarily originate from well-defined populations. There has recently been a drive to utilise alternative data sources, such as administrative data, which often include missing values that are not missing randomly -- but with certainty. Similarly, data sets arising from fusing or linking multiple data sources often include subjects who were excluded from at least one of the data sources, resulting in values missing with certainty in the linked data sets {(see \citealp{Gelman1998}, for an early approach for dealing with this)}. 
\par 
This notion of ``missing with certainty'' can be tied to inclusion probabilities. Whenever a subject has a zero inclusion probability, it will result in unit non-response that is non-random. Identifying these case{}{s} relies on knowledge relating to the construction of the data {set}, such as ascertaining how subjects were selected, what the underlying population was, and whether any data linkage took place.
\subsection{Dependencies between missingness indicators}
{The missingness indicator matrix $\mathbf{M}$ comprises just zeros and ones. As seen in Section 2, \citeauthor{Rubin1976}'s taxonomy implicitly assumes that the columns of this matrix -- which correspond to variables' missingness indicator vectors -- are (conditionally) independent: that is, missingness in one variable is (conditionally) independent of missingness in other variables. Specifically:}
\begin{align*}
\text{Under MCAR: \ } & M_j \independent M_k, & \forall \; j \neq k \quad \text{with} \quad j, k \in \{1, \ldots, p\}, \\
\text{Under MAR: \ } & M_j \independent M_k \mid \mathbf{X}_\text{obs}, & \forall \; j \neq k \quad \text{with} \quad j, k \in \{1, \ldots, p\}, \\
\text{Under MNAR: \ } & M_j \independent M_k \mid \mathbf{X}, & \forall \; j \neq k \quad \text{with} \quad j, k \in \{1, \ldots, p\},
\end{align*}
where $\independent$ denotes independence. Yet, in practice, these assumptions may be violated: there may be more complex dependencies between certain variables' missingness vectors. That is, there may be a multivariate structure to the matrix $\mathbf{M}$. \par  The notion of dependencies between the variables $M_1, \hdots, M_p$ adds a layer of complexity when modelling missingness mechanisms, giving scope for a range of complex interactions between $\mathbf{M}$ and $\mathbf{X}$. Essentially, rather than considering, say, $M_j$ as a function of $\mathbf{X}$ alone, we can consider $M_j$ to be a function of $\mathbf{X}$ \textit{and} $\mathbf{M}_{-j}$. \par To see how such dependencies arise in practice, let us consider the following example. Suppose batch failure occurs -- by a component failing, for example -- preventing a series of measurements from being collected. This would result in measurements prior to the failure being observed, while measurements post-failure would be missing. 
\section{A taxonomy for Structured Missingness} \label{sec4}
In this section, we set out our taxonomy, which, for a given variable $j$, hinges on the relationship between $M_j$, $\mathbf{M}_{-j}$, and $\mathbf{X}$. Existing missingness mechanisms focus on the relationship between $M_j$ and $\mathbf{X}$; for example, the taxonomy of MCAR, MAR, and MNAR broadly looks at whether $M_j$ depends on $\mathbf{X}$, $\mathbf{X}_\text{obs}$, or neither. By considering interactions between $M_j$ and $\mathbf{M}_{-j}$, too – that is, by considering that missingness in other variables can affect missingness in a variable – {a dimension of SM cases relating to multivariate missingness mechanisms is formed}. \par We continue to use the terms (i) MCAR,  (ii) MAR, and (iii) MNAR, for when: (i) $M_j$ is independent of $X$, (ii) $M_j$ depends on $\mathbf{X}_\text{obs}$, and (iii) $M_j$ depends on $\mathbf{X}$ ($\mathbf{X}_\text{mis}$ and $\mathbf{X}_\text{obs}$). We term the mechanisms where $M_1, \hdots, M_p$ are independent as the ``unstructured'' cases, and the mechanism where dependencies exist among $M_1, \hdots, M_p$ as the ``structured'' cases. \par In statistical modelling, and missing data methods more widely, it is generally assumed that the rows and columns (subjects and variables) of the data matrix -- and hence the missingness indicator matrix -- are interchangeable. However, many non-survey data sets have a natural, time-based ordering, as data are often collected in a sequential fashion. Therefore, while a variety of structural forms may be observed, we distinguish specifically between two forms of structure: \textbf{block structure} and \textbf{sequential structure}. The former relates to the case where missingness in a variable can be affected by any other variables; the latter relates to the case where missingness in a variable can only be affected by missingness in earlier variables.  \par When considering the missingness between a pair of variables, we can also distinguish between a \textbf{positive} and \textbf{negative} SM relationship. The former is the case where missingness in one variable \textit{increases} the probability of missingness in another variable. The latter is the case where missingness in one variable \textit{decreases} the probability of missingness in another variable. \par Finally, for each of the mechanisms we define, we give an example of how such a mechanism could arise in a clinical setting. We also use a {directed graph (DG)} to visualise the relationships associated with each mechanism, {not altogether dissimilar to the directed graphs (DGs) in \cite{mohan2021} for describing causal relationships in incomplete data.} 
\subsection{Cases relating to MCAR} \label{MCAR}
We begin by defining cases of SM that relate to MCAR, in the sense that there are no associations between $M_j$ and $\mathbf{X}$ for each $j \in \{1,\hdots,p\}$; that is, missingness depends on neither the observed nor the missing portion of the data. We do assume, however, that associations exist between the missingness indicators $M_1, \hdots, M_p$. We express mechanisms for ${M}_{j}$, therefore, in terms of $\mathbf{M}_{-j}$, and $\gamma$ (where $\gamma$ is used simply to represent the {}{parameters characterising the mechanism}). 
\subsubsection{MCAR -- Unstructured (MCAR-U)}
The MCAR-U mechanism is what is currently known as MCAR. It is the case where the missingness mechanism for $M_j$ ($j\in \{1, \hdots, p \})$ is independent of both the observed ($\mathbf{X}_\text{obs}$) and missing portions of the data ($\mathbf{X}_\text{mis}$), and also independent of missingness in other variables ($\mathbf{M}_{-j}$). 
\begin{definition} \normalfont\label{def4.1}
A missingness mechanism $\mathbf{M}=(M_1, \hdots, M_p)$ is MCAR-U if, for each variable $M_j$,
\begin{align*}
    p({M}_j\mid \mathbf{M}_{-j}, \mathbf{X}, \gamma) &= p({M}_j\mid\gamma) \quad \text{for all}\; \mathbf{X}, \gamma.
\end{align*} 
\end{definition}
MCAR-U missingness mechanisms have a tendency to display structure, even though there are no dependencies within $\mathbf{M}$ itself. Rearranging variables and subjects – in order of increasing number of missing values, for example – often creates the impression of structure. We call this \textbf{apparent structure}.   \par
The DG in Figure \ref{DG1} illustrates the relationship between the $X_{ij}$ and $M_{ij}$ in an MCAR-U mechanism. {This shows, as do all subsequent DGs that follow, three observations of a single variable, say, $X_{11}$, $X_{12}$, and $X_{13}$, along with their missingness indicators $M_{11}$, $M_{12}$, and $M_{13}$. The $X_{ij}$ and $M_{ij}$ are represented by blue squares and red diamonds, respectively; with solid red diamonds denoting the case where $X_{ij}$ is observed $(M_{ij}=0)$, and uncoloured red diamonds denoting the case where $X_{ij}$ is missing $(M_{ij}=1)$. The blue squares are always solid, representing the true underlying value. We always assume $X_{11}$ (the left blue square) is observed, and that $X_{12}$ and $X_{13}$ (the centre and right blue squares) are missing; thus it follows that the left red diamond is always coloured and the centre and right red diamonds are uncoloured. {When illustrating sequential structures, we assume, without loss of generality, that $X_{11}$, $X_{12}$, and $X_{13}$ are temporally ordered. Edges (for which there are none in this first instance) indicate relationships between the variables, with directed edges (arrows) indicating causal effects and bi-directed edges/arrows indicating associative effects. Dashed arrows indicate a probabilistic relationship and solid arrows indicate a deterministic relationship.}} This DG shows no association between the $X_{ij}$ and $M_{ij}$, because missingness is unstructured and MCAR. \par To give an example of how an MCAR-U mechanism can arise in a clinical setting, suppose the matrix $\mathbf{X}$ relates to laboratory test results; that is, suppose each row gives a particular subject's results, and each column gives the results for a particular test. Suppose some tests randomly fail owing to entirely technical reasons. Then, $M_{j}$, the missingness indicator vector for the $j$th variable ($j$th test), would be independent of both the test results themselves ($X$) and the missingness indicators of other variables ($\mathbf{M}_{-j}$).  
\begin{figure}[h!]
\centering
\includegraphics[width=7.2cm]{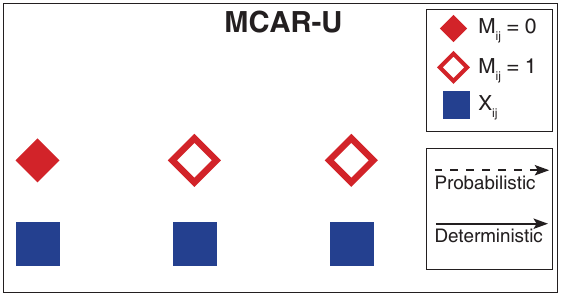}
\caption{\label{DG1} \textbf{MCAR-U}: A DG visualising the underlying relationship between the $X_{ij}$ and $M_{ij}$, which are represented by blue squares and red diamonds, respectively. In this instance, there is no relationship.}
\end{figure}         
\subsubsection{MCAR -- Weak Structure  (MCAR-WS)}
As we are considering cases of SM related to MCAR, we continue to assume no relationship between ${M}_{j}$ and $\mathbf{X}$ for each $j\in \{1, \hdots, p \}$. We now assume, however, that a relationship exists between ${M}_j$ and $\mathbf{M}_{-j}$. The MCAR-WS mechanism effectively assumes that missingness in {at least} one variable affects the probability of missingness in another variable. As an example, suppose that if $X_1$ is missing {{($M_1=1$)}}, the probability of missingness in $X_2$ {{($M_2=1$)}} is $q_1$; whereas, if {{$X_1$}} is observed {{($M_1=0$)}}, the probability of missingness in $X_2$ 
is $q_2$. {{Thus, there is a probabilistic relationship, which we term \textbf{weak structure}}}.
\begin{definition} \normalfont \label{def4.3}
A missingness mechanism $\mathbf{M}=(M_1, \hdots, M_p)$ is MCAR-WS if, for each variable $M_j$,
\begin{align*}
    p({M}_{j}\mid \mathbf{M}_{-j}, \mathbf{X}, \gamma) &= p({M}_j\mid\mathbf{M}_{-j}, \gamma) \quad \text{for all}\; \mathbf{X}, \gamma.
\end{align*} 
\end{definition}
Thus, unlike with MCAR-U, each ${M}_j$ {now} depends on $\mathbf{M}_{-j}$. \par
Definition \ref{def4.3} can be said to relate to block structure: it implicitly assumes that the columns in $\mathbf{M}$ do not have a given ordering. {A sequential structure relies on an ordering of the variables, $X_1, \hdots, X_p$, with missingness in ${X}_j$ only depending on $M_1, \hdots, M_{j-1}$, {and not on $M_{j+1}, \hdots, M_{p}$}; that is, {when missingness} depends only on variables appearing earlier in the sequence.} The sequential structure, then, can be viewed as a special case of block structure, when missingness in one variable depends only on certain variables and not all variables. There are similarities, too, between sequential missingness structures and missing data in longitudinal studies, which also have a given ordering {to the missingness} \citep{Laird1988, Diggle1994}. {For example, \citeauthor{Diggle1994}'s definitions of completely random drop-out (CRD), random drop-out (RD), and informative drop-out (ID), {}{could potentially have useful interpretations here.}}  \par For a practical example, again suppose the matrix $\mathbf{X}$ denotes subjects' test results, this time collected over several visits to the clinic. An MCAR-WS mechanism would occur if, once a subject misses a visit (say $M_j=1$), they are more likely to miss subsequent visits (the probabilities $p(M_{j+1}=1),\; p(M_{j+2}=1), \hdots$ increase). In this instance, a sequential structure would arise, but if there was no ordering to the variables, it would result in a block structure. \par {The left frame of Figure \ref{DG3} presents an MCAR-WS mechanism with a block structure (arrows point in both directions). The right frame presents a sequential structure (arrows point in one direction). {Crucially, the arrows are among the missingness indicators only.}}
\begin{figure}[h!]
\centering
\includegraphics[width=7.2cm]{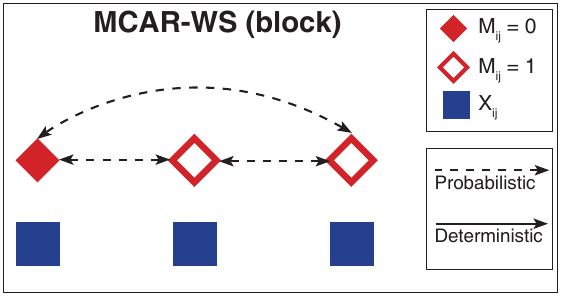}
\includegraphics[width=7.2cm]{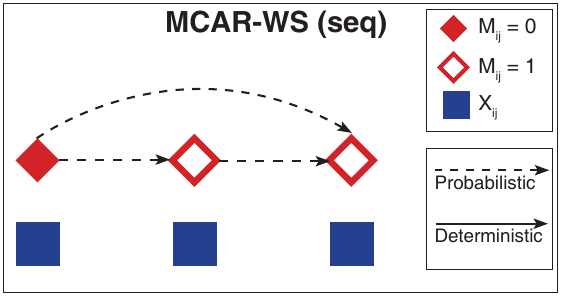}
\caption{\label{DG3} \textbf{MCAR-WS}: The left frame relates to a block structure. The right frame relates to a sequential structure.}
\end{figure} 
\subsubsection{MCAR -- Strong Structure  (MCAR-SS)}
The MCAR-SS mechanism is when missingness in one {or more} variables implies missingness in another variable with certainty. {{Revisiting our earlier example, where $q_1$ is the probability of missingness in $X_2$ given that $X_1$ is missing, \textbf{strong structure} is the case where $q_1$ is equal to 1}}. Hence, MCAR-SS can be viewed as a special case of MCAR-WS, where the relationship is no longer probabilistic but deterministic. \par {{Strong structure does not only include the case of a positive relationship in missingness across variables – that is, it is not just the case where missingness in one {or more} variables implies missingness in another – it also includes the case of a negative relationship: that is, the case where the observing of one or more variables implies missingness in another with certainty. }}  \par As before, since we are broadly dealing with MCAR, we continue to assume no relationship between ${M}_{j}$ and $\mathbf{X}$ for each $j\in \{1, \hdots, p \}$.
\begin{definition} \normalfont\label{def4.4}
A missingness mechanism $\mathbf{M}=(M_1, \hdots, M_p)$ is MCAR-SS if, for each variable $M_j$, there exists $i \in A \subseteq \{1, \hdots, n\}$, such that
\begin{align*}
    p({M}_{ij}=1\mid \mathbf{M}_{-j}, \mathbf{X}, \gamma) &= p({M}_{ij}=1\mid\mathbf{M}_{-j}, \gamma) =1 \quad \text{ for all}\;  \mathbf{X}, \gamma.
\end{align*} 
\end{definition}
Thus, there is an element of certainty with MCAR-SS that is not present in MCAR-WS. We distinguish between {block} and {sequential structure}, where the latter assumes an underlying ordering of the variables which the former does not have. \par To see the difference between MCAR-SS and MCAR-WS from a practical perspective, we revisit the previous example: {MCAR-SS would arise if, once a subject misses a visit ($M_j=1$), they are dropped from the study ($M_j=1 \Rightarrow M_{j+1}=1, M_{j+2}=1, \hdots$)}. The DGs in Figure \ref{DG4} also present the difference between MCAR-SS and MCAR-WS: the arrows of the DG are now solid -- not dashed -- which indicates certainty. 
\begin{figure}[h!] 
\centering
\includegraphics[width=7.2cm]{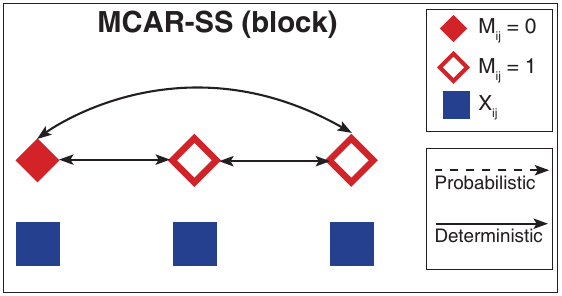}
\includegraphics[width=7.2cm]{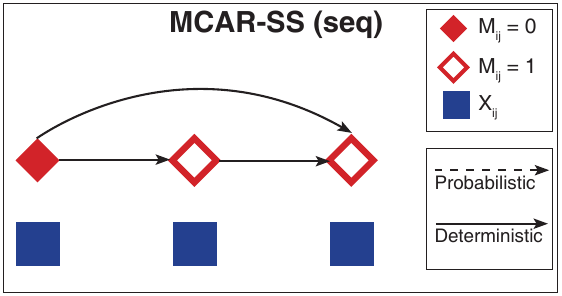}
\caption{\label{DG4} \textbf{MCAR-SS}: The left frame relates to a block structure, where causal arrows point in both directions, and the right frame relates to a sequential structure, where causal arrows point in {}{one direction only}. The arrows are now solid, which indicates a deterministic relationship: missingness in one variable implies missingness in another with certainty. }
\end{figure} 

{A key point to note here is that traditionally MCAR is viewed as a relatively simple, uninteresting scenario. However, what we show here is that, even with MCAR mechanisms, when considering the additional dimension of SM, the potential for a range of settings exists, varying in complexity.}

\subsection{Cases relating to MAR} \label{MAR}
We now move on to cases of SM that relate to MAR. The setup {}{will resemble that of the previous section}, but the difference is that we now consider interactions between $\mathbf{X}_\text{obs}$ (the observed portion of the data) and $\mathbf{M}$. Throughout this {}{section}, therefore, we will express mechanisms in terms of $\mathbf{M}_{-j}$, $\mathbf{X}_\text{obs}$, and $\gamma$. {An important point with these MAR cases is that ${X}_{ij}$ cannot have a direct effect on its corresponding ${M}_{ij}$, as this would relate to MNAR.} 
\subsubsection{MAR -- Unstructured, Probabilistic  (MAR-UP)}
As with the MCAR cases of SM, we begin with the unstructured MAR mechanisms, where {$M_j$} depends on the observed data $\mathbf{X}_\text{obs}$, but does not depend on missingness in the other variables $\mathbf{M}_{-j}$. \par We first consider the case of a \textbf{probabilistic} mechanism – the notion of which is similar to that of weak structure from MCAR-WS – which is when $\mathbf{X}_\text{obs}$ affects the \textit{probability} that ${M}_{ij}=1$. 
\begin{definition} \normalfont\label{def5.1}
A missingness mechanism $\mathbf{M}=(M_1, \hdots, M_p)$ is MAR-UP if, for each variable $M_j$,
\begin{align*}
    p({M}_j\mid \mathbf{M}_{-j}, \mathbf{X}, \gamma) &= p({M}_j\mid \mathbf{X}_\text{obs},\gamma) \quad \text{for all}\; \mathbf{X}_\text{obs}, \gamma.
\end{align*} 
\end{definition}
The MAR-UP definition is identical to the current definition of MAR, and, although unstructured, can display apparent structure, especially if a similar set of variables is influencing missingness in each variable. \par MAR-UP may arise, for example, if physicians are less inclined (without there being consistency of decisions) to give a particular test $X_j$ to elderly patients. {In this way, $M_j$ ($X_j$'s missingness indicator vector) depends only {on} patients' ages ($\mathbf{X}_\text{obs}$) and not on missingness in other variables $(\mathbf{M}_{-j})$.}
\subsubsection{MAR -- Unstructured, Deterministic  (MAR-UD)}
Similarly, in the case of \textbf{deterministic} unstructured MAR mechanisms, $\mathbf{X}_\text{obs}$ directly dictates whether ${M}_{ij}=1$. Thus, it is a special case of MAR-UP where values are missing with certainty. 
\begin{definition} \normalfont\label{def5.2}
A missingness mechanism $\mathbf{M}=(M_1, \hdots, M_p)$ is MAR-UD if, for each variable $M_j$, there exists $i \in A \subseteq \{1, \hdots, n\}$, such that
\begin{align*}
    p({M}_{ij}=1\mid \mathbf{M}_{-j}, \mathbf{X}, \gamma) &= p({M}_{ij}=1\mid \mathbf{X}_\text{obs},\gamma) = 1  \quad \text{for all}\; \mathbf{X}_\text{obs}, \gamma.
\end{align*} 
\end{definition}
The MAR-UD mechanism would arise, for example, if physicians \textit{never} give a particular test to elderly patients. {Now, patients' ages ($\mathbf{X}_\text{obs}$) have a deterministic effect on $M_j$.}  \par 
The DGs in Figure \ref{DG5} illustrate the basic structure of the MAR-UP and MAR-UD mechanisms. In both instances, there is now {an effect {of $X_{11}$ (left blue square), which is observed, on $M_{12}$ and $M_{13}$} (uncoloured red diamonds)}; with MAR-UP the arrows are dashed, and with MAR-UD the arrows are solid.
\begin{figure}[h!]
\centering
\includegraphics[width=7.2cm]{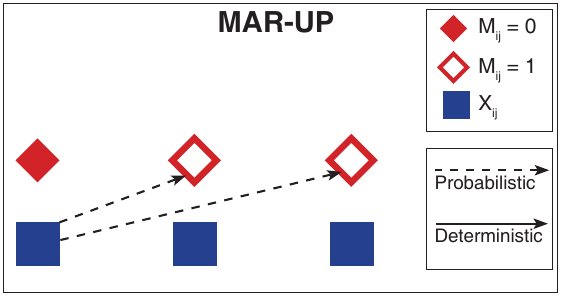}
\includegraphics[width=7.2cm]{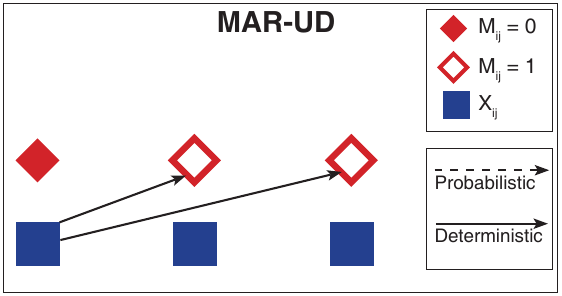}
\caption{\label{DG5} \textbf{MAR-UP} and \textbf{MAR-UD}: The left frame gives the MAR-UP mechanism; the right frame gives the MAR-UD mechanism.}
\end{figure} 
\subsubsection{MAR -- Weak Structure (MAR-WS)}
We now {}{additionally} assume that relationships exist between ${M}_j$ and $\mathbf{M}_{-j}$, {}{\textit{as well as}} between ${M}_j$ and $\mathbf{X}_\text{obs}$. As before, we distinguish between a weak structure, when $\mathbf{M}_{-j}$ and $\mathbf{X}_\text{obs}$ taken together affect the probability distribution of ${M}_j$, and strong structure, when $\mathbf{M}_{-j}$ and $\mathbf{X}_\text{obs}$ directly determine whether ${M}_{ij}=1$. Once again, we also distinguish between {block} and {sequential structure}, where the latter assumes an underlying ordering to the variables and missingness depends only on variables {}{earlier in the sequence}. 
\begin{definition} \normalfont\label{def5.3}
A missingness mechanism $\mathbf{M}=(M_1, \hdots, M_p)$ is MAR-WS if, for each variable $M_j$,
\begin{align*}
    p({M}_{j}\mid \mathbf{M}_{-j}, \mathbf{X}, \gamma) &= p({M}_j\mid\mathbf{M}_{-j}, \mathbf{X}_\text{obs}, \gamma) \quad \text{for all}\; \mathbf{X}_\text{obs}, \gamma.
\end{align*} 
\end{definition}
MAR-WS may arise if physicians do not always give a particular test $(X_j)$ to elderly patients – thus $\mathbf{X}_\text{obs}$ has an effect on $M_j$ – and which in turn then means they are less likely to be invited back for further tests ($M_j=1$ increases the probabilities $p(M_{j+1}=1),\;p(M_{j+2}=1), \hdots$). \par {The left and right frames of Figure \ref{DG6} present a MAR-WS mechanism with a block (causal arrows point in both directions) and sequential structure (causal arrows point in one direction), respectively. As with the DG for MAR-UP (Figure \ref{DG5}), there is a probabilistic causal arrow from $X_{11}$ (observed; left blue square) to $M_{12}$ and $M_{13}$ (uncoloured red diamonds); and, in addition, there are now {relationships} {between $M_{11}$, $M_{12}$, and $M_{13}$.}}
\begin{figure}[h!]
\centering
\includegraphics[width=7.2cm]{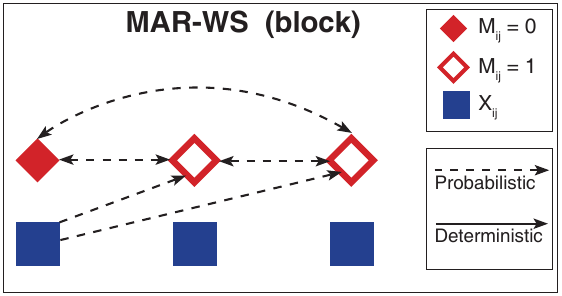}
\includegraphics[width=7.2cm]{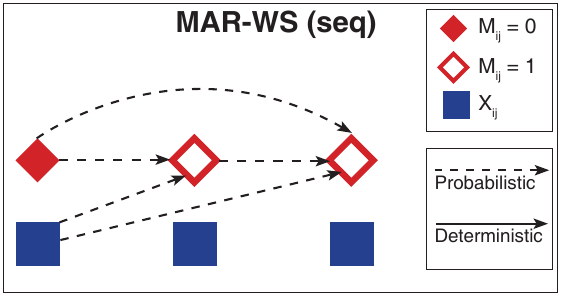}
\caption{\label{DG6} \textbf{MAR-WS}: The MAR-WS mechanism with a block structure (left frame) and a sequential structure (right frame).}
\end{figure} 
\subsubsection{MAR -- Strong Structure (MAR-SS)}
The MAR-SS mechanism is the case where $\mathbf{X}_\text{obs}$ and $\mathbf{M}_{-j}$, taken together, directly dictate whether ${M}_{ij}=1$. Thus, it is the deterministic case of MAR-WS. 
\begin{definition} \normalfont\label{def5.6}
A missingness mechanism $\mathbf{M}=(M_1, \hdots, M_p)$ is MAR-SS if, for each variable $M_j$, there exists $i \in A \subseteq \{1, \hdots, n\}$, such that
\begin{align*}
    p({M}_{ij}=1\mid \mathbf{M}_{-j}, \mathbf{X}, \gamma) &= p({M}_{ij}=1\mid\mathbf{M}_{-j}, \mathbf{X}_\text{obs}, \gamma)=1 \quad \text{ for all}\;  \mathbf{X}_\text{obs}, \gamma.
\end{align*} 
\end{definition}
To see how the mechanism MAR-SS can arise, we reconsider the previous example. Now suppose tests are \textit{only} granted to subjects below a certain age {(where age is observed in $\mathbf{X}_\text{obs}$)}, and that once a subject misses one test ($M_j=1$), they are automatically dropped from a study ($M_j=1 \Rightarrow M_{j+1}=1, M_{j+2}=1, \hdots$).    \par
{The left and right frames of Figure \ref{DG9} give a MAR-SS mechanism with a block (arrows between $M_{11}$, {$M_{12}$}, and $M_{13}$ point in both directions) and sequential structure (one directional arrows), respectively. The difference compared with the MAR-WS case (Figure \ref{DG9}) is that {the} arrows are {now} solid.}
\begin{figure}[h!]
\centering
\includegraphics[width=7.2cm]{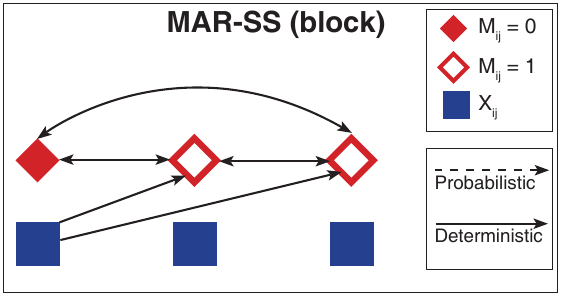}
\includegraphics[width=7.2cm]{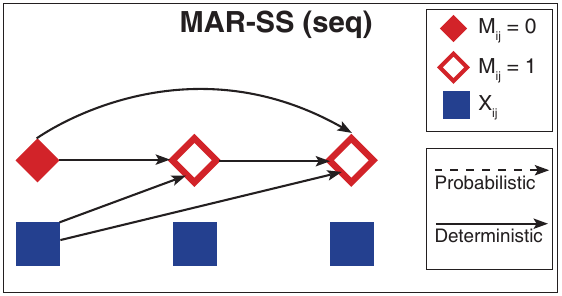}
\caption{\label{DG9} \textbf{MAR-SS}: The left frame gives a block structure; the right frame gives a sequential structure. There is a deterministic causal effect from $\mathbf{X}_\text{obs}$ to ${M}_{ij}$ and a strong structure between the ${M}_{ij}$. }
\end{figure} 
\subsection{Cases relating to MNAR}
Lastly, we move on to cases of SM that relate to MNAR. {We now assume that the ${X}_{ij}$ do, indeed, have a direct effect on the corresponding ${M}_{ij}$. In a similar way in which Section \ref{MAR} builds on Section \ref{MCAR}, this section builds on Section \ref{MAR}, with the difference being that mechanisms now {}{additionally include} the missing portion of the data $\mathbf{X}_\text{mis}$ in their expressions; that is, we express mechanisms in terms of $\mathbf{M}_{-j}$, $\mathbf{X}_\text{obs}$, $\mathbf{X}_\text{mis}$, and $\gamma$.} 
\subsubsection{MNAR -- Unstructured, Probabilistic  (MNAR-UP)}
In this probabilistic instance, which is identical to the current definition of MNAR, $\mathbf{X}_\text{obs}$ and $\mathbf{X}_\text{mis}$ (or more simply, $\mathbf{X}$) affect the {probability} that ${M}_{ij}=1$. 
\begin{definition} \normalfont\label{def6.1}
A missingness mechanism $\mathbf{M}=(M_1, \hdots, M_p)$ is MNAR-UP if, for each variable $M_j$,
\begin{align*}
    p({M}_j\mid \mathbf{M}_{-j}, \mathbf{X}, \gamma) &= p({M}_j\mid \mathbf{X}_\text{obs},\mathbf{X}_\text{mis},\gamma) \quad \text{for all}\;  \mathbf{X}_\text{obs}, \mathbf{X}_\text{mis}, \gamma.
\end{align*} 
\end{definition}
{To give an example of how MNAR-UP may arise, suppose $X_j$ gives peak expiratory flow (PEF) measurements, a test commonly given to patients with asthma. A particularly bad case of asthma – a case which, if the test is carried out, would return a low PEF result $(\mathbf{X}_\text{mis})$ – may preclude even taking a PEF test (for example, the patient may be too ill to visit the clinic), thus causing a missing test result. }
\subsubsection{MNAR -- Unstructured, Deterministic  (MNAR-UD)}
In the {deterministic} instance of MNAR-UD, $\mathbf{X}_\text{obs}$ and $\mathbf{X}_\text{mis}$ directly dictate whether ${M}_{ij}=1$.
\begin{definition} \normalfont\label{def6.2}
A missingness mechanism $\mathbf{M}=(M_1, \hdots, M_p)$ is MNAR-UD if, for each variable $M_j$, there exists $i \in A \subseteq \{1, \hdots, n\}$, such that
\begin{align*}
    p({M}_{ij}=1\mid \mathbf{M}_{-j}, \mathbf{X}, \gamma) &= p({M}_{ij}=1\mid \mathbf{X}_\text{obs}, \mathbf{X}_\text{mis}, \gamma) =1 \quad \text{for all}\;  \mathbf{X}_\text{obs},\mathbf{X}_\text{mis}, \gamma.
\end{align*} 
\end{definition}
{{Continuing} with the asthma example, MNAR-UD can arise, for example, if PEF measurements falling below a certain value are incorrectly recorded as being missing.} \par
The DGs in Figure \ref{DG11} illustrate the basic structure of the MNAR-UP and MNAR-UD mechanisms. {In both instances, we now also have causal effects from $X_{12}$ and $X_{13}$ (centre and right blue squares) to their own missingness indicators $M_{12}$ and $M_{13}$ (uncoloured red diamonds). In the left and right frames, we have a block and sequential structure, respectively.}
\begin{figure}[h!]
\centering
\includegraphics[width=7.2cm]{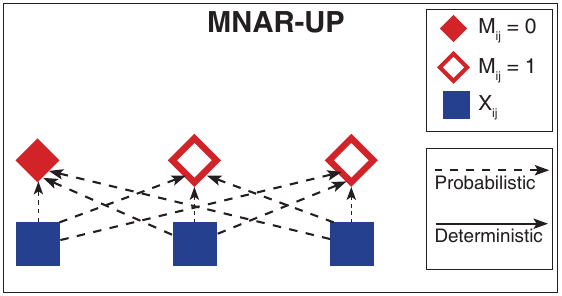}
\includegraphics[width=7.2cm]{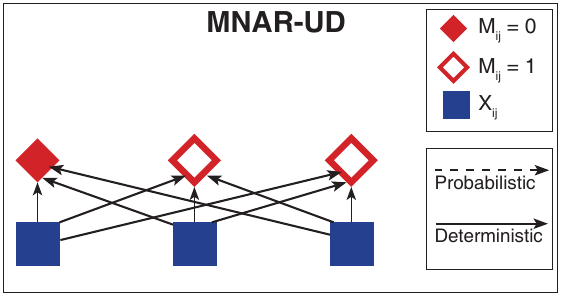}
\caption{\label{DG11} \textbf{MNAR-UP} and \textbf{MNAR-UD}: The left frame gives the MNAR-UP mechanism; the right frame gives the MNAR-UD mechanism.}
\end{figure} 
\subsubsection{MNAR -- Weak Structure (MNAR-WS)}
Again, we distinguish between weak and strong structure. In this instance, weak structure is when ${M}_j$ depends on both $\mathbf{M}_{-j}$ and $\mathbf{X}$.
\begin{definition} \normalfont\label{def6.3}
A missingness mechanism $\mathbf{M}=(M_1, \hdots, M_p)$ is MNAR-WS if, for each variable $M_j$,
\begin{align*}
    p({M}_{j}\mid \mathbf{M}_{-j}, \mathbf{X}, \gamma) &= p({M}_j\mid\mathbf{M}_{-j}, \mathbf{X}_\text{obs},\mathbf{X}_\text{mis}, \gamma) \quad \text{for all}\;   \mathbf{X}_\text{obs},\mathbf{X}_\text{mis}, \gamma.
\end{align*} 
\end{definition}
 {For the asthma example, the mechanism MNAR-WS can arise if, once a patient misses a visit to the clinic due to a severe case of asthma – an occasion on which a low PEF result $(\mathbf{X}_\text{mis})$ would have been returned had they attended – they begin to lose contact with the clinic (they become disengaged), leading to a lower likelihood of further testing (increasing the probabilities $p(M_{j+1}=1),\;p(M_{j+2}=1), \hdots$). } \par The DGs in Figure \ref{DG12} present the structure of an MNAR-WS mechanism. {As with MNAR-UP and MNAR-UD (Figure \ref{DG11}), we have causal effects from $X_{12}$ and $X_{13}$ (centre and right blue squares) to their own missingness indicators $M_{12}$ and $M_{13}$ (uncoloured red diamonds). In addition, we have {relationships} among the missingness indicators $M_{11}$, $M_{12}$, and $M_{13}$ (red diamonds). {In the left and right frames, the arrows between the missingness indicators are bi- and one-directional, representing block and sequential structures, respectively.}} 
\begin{figure}
\centering
\includegraphics[width=7.2cm]{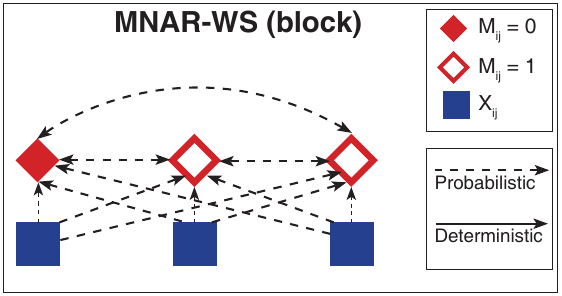}
\includegraphics[width=7.2cm]{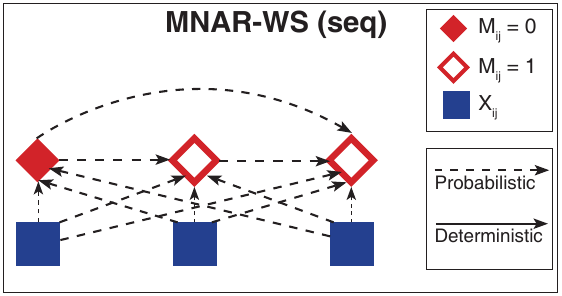}
\caption{\label{DG12} \textbf{MNAR-WS}: The left frame gives a block structure; the right frame gives a sequential structure. The missing data values are now affecting the missingness.}
\end{figure} 
\subsubsection{MNAR -- Strong Structure (MNAR-SS)}
Finally, the MNAR-SS mechanism is the case where both $\mathbf{M}_{-j}$ and $\mathbf{X}$ directly dictate whether ${M}_{ij}=1$; it is a special case of the weak structure. 
\begin{definition} \normalfont\label{def6.6}
A missingness mechanism $\mathbf{M}=(M_1, \hdots, M_p)$ is MNAR-SS if, for each variable $M_j$, there exists $i \in A \subseteq \{1, \hdots, n\}$, such that
\begin{align*}
    p({M}_{ij}=1\mid \mathbf{M}_{-j}, \mathbf{X}, \gamma) &= p({M}_{ij}=1\mid\mathbf{M}_{-j}, \mathbf{X}_\text{obs}, \mathbf{X}_\text{mis}, \gamma) =1 \quad \text{ for all}\; j, \mathbf{X}_\text{obs}, \mathbf{X}_\text{mis}, \gamma.
\end{align*} 
\end{definition}
{The mechanism MNAR-SS can arise if a severe case of asthma $(\mathbf{X}_\text{mis})$ automatically precludes taking a PEF test $(M_j=1)$, and which, in turn, automatically precludes all further testing ($M_j=1 \Rightarrow M_{j+1}=1, M_{j+2}=1, \hdots$).}    \par
{The DGs in Figure \ref{DG13} give a MNAR-SS mechanism. The difference compared to MNAR-WS (Figure \ref{DG12}) is that arrows are now {solid.}}
\begin{figure}[h!]
\centering
\includegraphics[width=7.2cm]{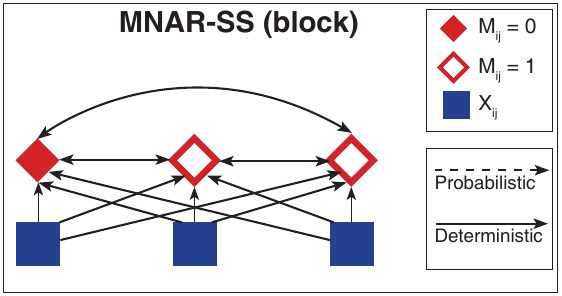}
\includegraphics[width=7.2cm]{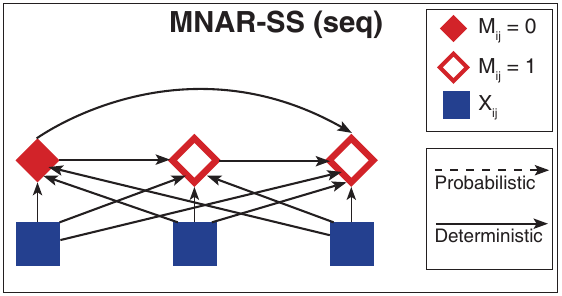}
\caption{\label{DG13} \textbf{MNAR-SS}: The left frame gives a block structure; the right frame gives a sequential structure. There is a deterministic causal effect from ${X}_{ij}$ to ${M}_{ij}$ and a strong structure between the ${M}_{ij}$.}
\end{figure} 
\subsection{Summary of SM cases}
{So far, we have built on \citeauthor{Rubin1976}'s taxonomy to describe and characterise a range of SM mechanisms (summarised in Figure \ref{family}) that vary according to the following dimensions:}
\begin{enumerate}
    \item The relationship between $M_j$ (for all $j\in \{1, \hdots, p \}$) and the observed and missing portion of the data, $\mathbf{X}_\text{obs}$ and $\mathbf{X}_\text{mis}$, which broadly relates to the concepts of MCAR, MAR, and MNAR. 
    \item Whether missingness is unstructured or structured; that is, whether $M_j$ depends on $\mathbf{M}_{-j}$.
    \item Whether the multivariate structure of missingness is weak or strong (probabilistic or deterministic); that is, whether $\mathbf{M}_{-j}$ influences or directly determines $M_j$. 
    \item Whether missingness occurs in either a block or sequential structure; that is, whether only previously observed variables $M_1, \hdots, M_{j-1}$ affect $M_j$ (given an underlying ordering to the variables).
\end{enumerate}
\begin{figure}[h!]
\centering
\includegraphics[width=14cm]{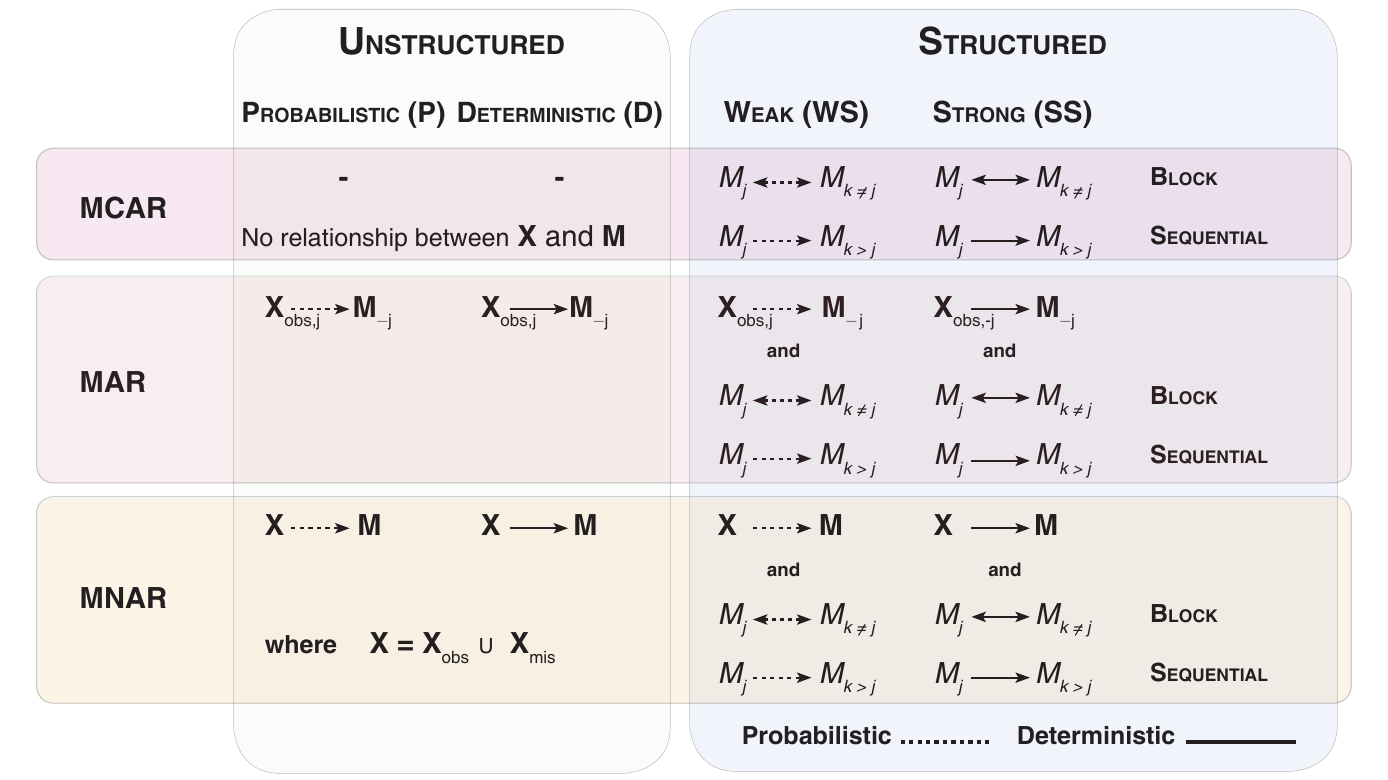}
\caption{\label{family} The range of {}{structured} missingness mechanisms considered in Section \ref{sec4}.}
\end{figure} 
\section{Structured Missingness: Other cases} \label{sec7}
While the definitions in Section \ref{sec4} cover the possible interactions between $M_j$, $\mathbf{M}_{-j}$, and $\mathbf{X}$, they do not exhaustively cover {all} cases of SM. In this section, we consider alternative ways in which structure can manifest in the missingness indicator matrix.
\subsection{The presence of a subject effect}
The missing data literature typically assumes that two subjects with the same characteristics have the same probability of missingness for a particular variable. Yet, in practice, this is unlikely to be the case; for example, some subjects are inherently more inclined to attend a clinic than others. This heterogeneity (one of the routes to SM considered by \citealp{Mitra2023}) can be accounted for via the presence of a {subject effect}, which can be conceptualised mathematically through a random effect $S_i$ $(i\in \{1, \hdots, n \})$ unique to individual $i$ and unrelated to any variables observed in the data.  \par As an example, in the presence of a subject effect, Definition \ref{def4.1} for MCAR-U can be amended to:
\begin{definition} \normalfont\label{def4.2}
A missingness mechanism $\mathbf{M}=(M_1, \hdots, M_p)$ is MCAR-U with a subject effect if, for each variable $M_j$,
\begin{align*}
    p({M}_{ij}=1\mid \mathbf{M}_{-j}, \mathbf{X}, S_i, \gamma) &= p({M}_{ij}\mid S_i,\gamma) \quad \text{for all}\; \mathbf{X}, S_i, \gamma.
\end{align*} 
\end{definition}
Note, we could add a subject effect to any of the mechanisms considered so far. \par The DG corresponding to MCAR-U with a subject effect is presented in Figure \ref{DG19}. The subject effect is denoted by the green circle. \par To add another layer of complexity, we could include dependencies between the $S_i$. Dependencies could arise, for instance, through clustering. In a cluster-randomised controlled trial, for example, children from the same school class, or subjects from the same region in a multi-centre trial, may have correlated random effects.
\begin{figure}[h!]
\centering
\includegraphics[width=7.2cm]{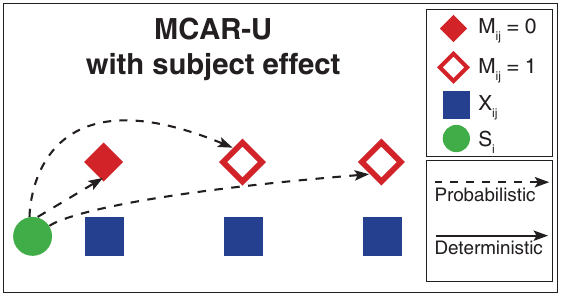}
\caption{\label{DG19} \textbf{MCAR-U with a subject effect}: There is now a green circle, denoting the subject effect, that is affecting missingness. }
\end{figure} 
 \subsection{Logical Missingness} \label{strucmiss}
There are also cases of missingness that are reminiscent of structural zeros in the categorical data literature, where counts in contingency tables comprise of two sorts: random (or sampling) zeros and structural zeros (see, for example, \citealp{Bishop1975, Agresti2012}). The former are zeros that arise through the random nature of sampling; if a similar sample is taken, such counts may not be necessarily zero. The latter are zero counts for which there is a logical reason why they must be zero. \par A similar concept holds in relation to missingness. We can further decompose cases into those for which there is a logical or biological reason why missingness occurred; that is, missingness {may} occur because it is fundamentally not possible for an entry in the data matrix to be observed; for example, questions relating to pregnancy are only relevant to women.  \par Logical missingness is unlike other cases of missingness, where underlying data values exist and make sense, but where the uncertainty of the data collection process -- as well as the uncertainty in the data values themselves -- 
{}{result in particular values being unobserved}. For example, a protocol may specify that only the most severe patients receive a certain test due to its invasive nature. By contrast, cases of logical missingness are when there is a fundamental reason for the missingness. For example, a prostate specific antigen test would not be {}{applicable to} females {}{resulting in these test results being {\em logically missing}}. \par In general, when faced with logical missingness, we can either ignore or weight out the values which are logically missing. In this instance, we would effectively be assuming that a missing mechanism is composed of (at least) two underlying mechanisms, which leads us on to the notion of multiple mechanisms. 
 \subsection{Multiple SM mechanisms} \label{com}
 {Multiple SM mechanisms are likely to be the norm in data sets developed at scale. Consider the case, for example, where there is a probabilistic relationship between the missingness indicators but a deterministic relationship with the data. For example, suppose $\mathbf{M}_{-j}$ affects the probability that ${M}_{ij}=1$, but $\mathbf{X}_\text{obs}$ dictates, in a deterministic fashion, whether ${M}_{ij}=1$.} This is essentially the union between the MCAR-WS and the MAR-SS mechanisms: $\mathbf{X}_\text{obs}$ first dictates whether certain values will be missing; and then these missing values influence missingness in other variables. An example of this in a clinical setting is when tests are only performed at clinic visits and only subjects above a certain age are invited to regular visits, but where other subjects may receive a test when they attend the clinic for other reasons. The DG in Figure \ref{DG20} shows how in this instance there is a weak structure (dashed line) between the ${M}_{ij}$ but  a strong structure between $\mathbf{X}_\text{obs}$ and the ${M}_{ij}$. \par It is straightforward to see how further similar mechanisms can arise through combining multiple mechanisms. For example, if $\mathbf{X}_\text{obs}$ only influences whether certain values will be missing, but then these missing values dictate whether certain values will be missing in other variables, we would have the union between the MAR-WS and the MCAR-SS mechanisms. 
 \begin{figure}[h!]
\centering
\includegraphics[width=7.2cm]{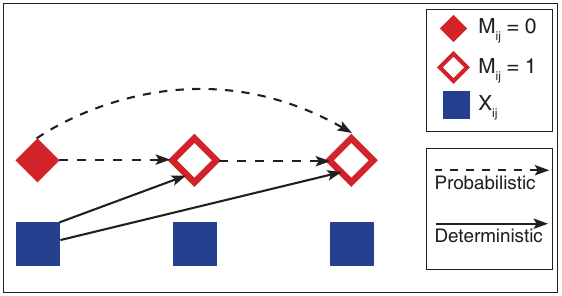}
\caption{\label{DG20} There is a weak structure (dashed line) between the ${M}_{ij}$ but  a strong structure between $\mathbf{X}_\text{obs}$ and the ${M}_{ij}$.}
\end{figure} 
\section{{Empirical illustrations of the impact of SM on inference and prediction}} \label{sec8}
We have introduced new definitions to cover the cases where the missingness mechanism for $M_j$ depends on $\mathbf{M}_{-j}$, yet a glaring question remains: \textit{does SM really matter?} {That is, does existing methodology, and statistical software, already satisfactorily deal with SM?} \par In this section, we demonstrate, through simulation examples, the impact of SM on inferential validity and prediction performance, showing that strong structures, in particular, pose unique challenges, {but also sometimes unique opportunities to exploit}. We consider three simulation examples, looking at structure in relation to MCAR, MAR, and MNAR, respectively. This section is by no means intended to be a comprehensive guide to dealing with SM, which is beyond the scope of this paper. Rather, its intention is to highlight the impact of SM from a practical perspective. 
\subsection{Simulation 1: Prediction in the presence of structured and unstructured MCAR mechanisms}
When assessing the impact of SM on the performance of statistical methods, there are two broad aspects to consider: predictive and inferential performance. In this first simulation, we highlight the impact of SM on prediction when considering structured and unstructured MCAR examples. Specifically, we consider the effect of various factors relating to missingness structure on prediction, including:
\begin{enumerate}
    \item Strength of missingness structure (strong structure vs. weak structure);
    \item Sequential and block structures;
     \item Missingness within the test data set;
        \item Correlation within the data matrix $\mathbf{X}$.
\end{enumerate}
We first generate simulated data sets with $p=10$ variables and $n=1100$ subjects – 100 rows are used for training; 1000 rows are used as the test data set – distributed according to a multivariate normal (MVN) distribution: 
\begin{align}
 \mathbf{X} &\sim \text{MVN}_{10}(\boldsymbol\mu, \boldsymbol\Sigma) \nonumber \\
 \text{where} \quad \boldsymbol\mu&=\begin{bmatrix}
    0 \\
   0  \\
    \vdots \\
    0 
\end{bmatrix} \quad \text{and} \quad \boldsymbol\Sigma = \begin{bmatrix}
    1 & \rho &  \dots  & \rho \\
   \rho & 1 & \dots  & \rho \\
    \vdots & \vdots & \ddots & \vdots \\
    \rho & \rho & \dots  & 1
\end{bmatrix}. \nonumber 
  \end{align}
For the parameter $\rho$, which defines the level of correlation within the data matrix $\mathbf{X}$, we consider two values: $\rho=0$, which equates to assuming the variables are independent; and $\rho=0.4$, which equates to assuming equal, non-zero covariances between all pairs of variables. \par {We next impose a range of unstructured and structured missingness mechanisms on $\mathbf{X}$. These include four MCAR-U mechanisms, labelled MCAR-U (1)–(4), which have different missingness rates across the $p=10$ variables. These are:
\begin{itemize}
    \item MCAR-U (1): Variables 1–10 have 45\% missingness. 
\item MCAR-U (2): Variable 1 has 0\% missingness, variable 2 has 10\% missingness, dvariable 3 has 30\% missingness, ..., and variable 10 has 90\% missingness.
\item MCAR-U (3): Variables 1–5 have 0\% missingness; variables 6–10 have 90\% missingness.
\item MCAR-U (4): Variable 1 has 0\% missingness; variables 2–10 have 50\% missingness.
\end{itemize}
We also consider MCAR-WS and MCAR-SS mechanisms (with both block and sequential structures); for comparison, we consider a complete data set and a case of block missingness (block missingness in the sense that we fully observe some subjects and never observe the others). For every structure except the complete case, there is approximately 45\% missingness, and therefore the only difference between the mechanisms is in the distribution of the missing values, that is, the missing data pattern. The various missingness structures are listed and shown visually in Figure \ref{missingnessplots}, with blue tiles representing observed values and orange and grey tiles representing missing values of $\mathbf{X}$ (we go on to impute the orange tiles and delete the grey tiles).} \par
 \begin{figure}[h!]
\centering
        \includegraphics[width=14.5cm]{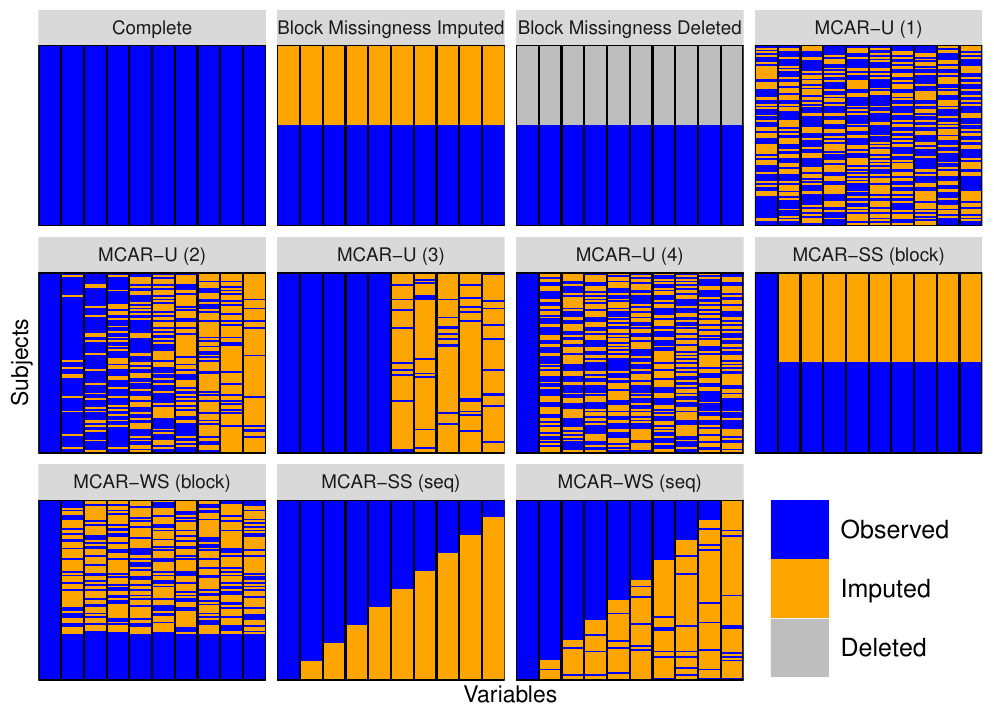}
        \caption{\label{missingnessplots} The different missingness structures compared in the first simulation. } 
\end{figure} 
We multiply impute missing values using fully conditional specification (FCS), an approach which has various names, including Sequential Regression Multiple Imputation (SRMI) \citep{Raghunathan2001} and Multivariate Imputation via Chained Equations (MICE) \citep{vanBuuren2011}, and which is implemented in the R package \pkg{mice}. We use the default settings, which includes the use of predictive mean matching (\texttt{method=`pmm'}), 5 iterations of the Gibbs sampler (\texttt{maxit=5}), and the generation of $m=5$ imputed data sets (\texttt{m=5}). We then use these imputation models to multiply impute missing values in our test set, too; this is achieved through the \texttt{ignore} argument in \pkg{mice}.  \par We consider the quantity $Y$, linked to $\mathbf{X}$ via the following analysis (substantive) model,
\begin{align*}
    Y= \alpha_0 + \sum\limits_{i=1}^{10}\alpha_i X_i + \varepsilon_i \quad \text{where} \quad \varepsilon \sim N(0,\sigma^2),
\end{align*}
and where $\sigma^2=4$. We use our training data to estimate: (i) the imputation model parameters, and (ii) the analysis model's parameters $\boldsymbol\alpha=(\alpha_0, \hdots, \alpha_{10})$, which we then use to predict $Y$ in our test set. The true values for $\boldsymbol\alpha$ are $\alpha_i=1$ for all $i$. We can then compare the predictions for $Y$ with the true values, and compute the mean squared error (MSE) for our test set {as a measure of predictive performance}. The total error observed can be decomposed into two sources, error due to inaccuracies in the model fit, that is, the estimation of $\boldsymbol\alpha$, and error due to inaccuracies in imputing values into the test data set. To distinguish between these two sources of error, we also run the simulation with no missing values in the test set. To briefly summarise, therefore, there are three factors at work in the simulation:
\begin{enumerate}
    \item The correlation between the variables in $\mathbf{X}$ (either $\rho=0$ or $\rho=0.4$).
    \item The type of MCAR missingness mechanism imposed on $\mathbf{X}$ (11 examples considered; see Figure \ref{missingnessplots}).
     \item Whether the test set is complete or includes missing values (complete or missing).
\end{enumerate}
This returns $2\times 11 \times 2=44$ combinations of factors. The MSEs for these combinations, over $n_\text{sim}=1000$ simulation runs, are presented in the boxplots in Figure \ref{sim1MSE}. The results for when the test set is (i) complete or (ii) includes missing values are given in the red and blue boxplots, respectively. \par From a general perspective, the structured mechanisms return greater error than the unstructured mechanisms when the variables in $\mathbf{X}$ are correlated rather than uncorrelated (the bottom four rows vs. the top seven rows). This result is far from obvious: typically for MCAR mechanisms, values in $\mathbf{X}$ are seen to neither influence nor be influenced by the missing values. Yet here, the missing data pattern is clearly affecting the results of an analysis performed on $\mathbf{X}$. \par In a similar way, for the structured mechanisms there is greater error when the variables in $\mathbf{X}$ are correlated rather than uncorrelated (right vs. left boxplots). This too is not obvious for MCAR mechanisms, where $\mathbf{X}$ and $\mathbf{M}$ are independent. \par 
The MCAR-SS (block) mechanism, when we impute missing values in the test set (blue boxplot) and when $\rho=0.4$, results in the greatest overall error. By contrast, however, the other three boxplots relating to MCAR-SS (block) show relatively small amounts of error. This illustrates the danger of strong structures, in the sense that key relationships between variables can easily be lost. The MSE metric can be expressed as squared bias plus variance; in this instance, as expected, the error is arising through higher variances rather than bias. \par 
The overriding point from this simulation is that missingness structure clearly has an impact on predictive ability – and this relationship is not obvious from the outset. More generally, the effect of missingness mechanisms on predictive ability is an underexplored area of research (see \citealp{hornung2023}) for further investigation in this area)}, especially the effect of different {types of SM}.
  \begin{figure}[h!]
\centering
        \includegraphics[width=17cm]{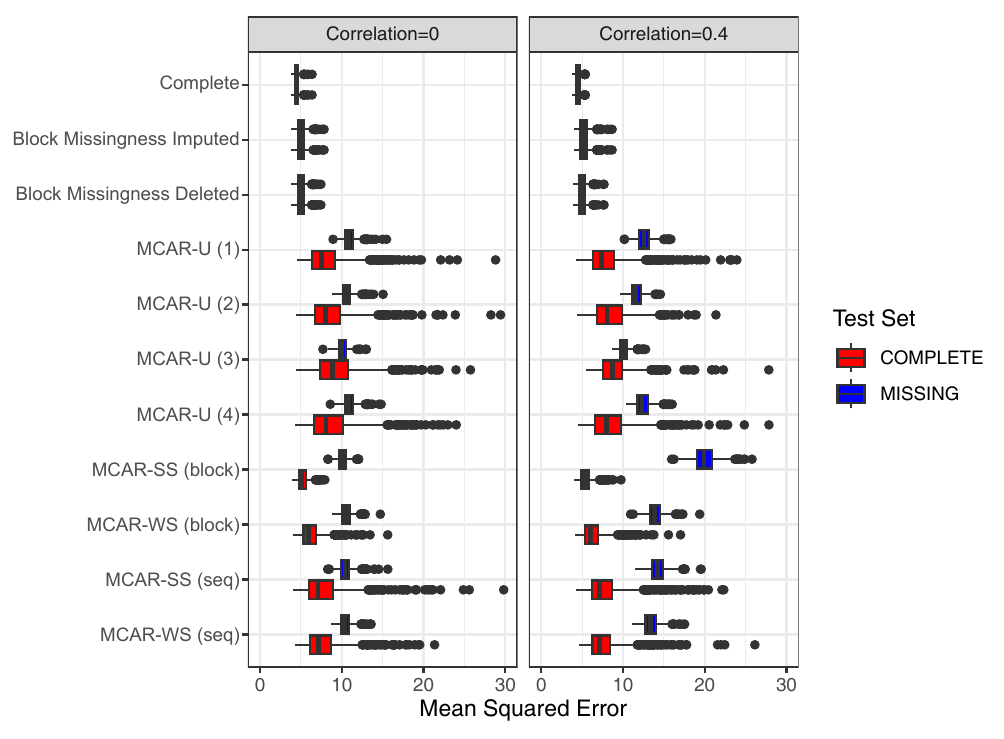}
        \caption{\label{sim1MSE} Boxplots showing the MSE values for various missingness structures. The left- and right-hand panels give the results for when the variables in $\mathbf{X}$ are independent ($\rho=0$), and for when $\rho=0.4$, respectively. The red boxplots give the results for when {the test set is complete, and the blue boxplots give the results for when missing values occur in the test set}.} 
\end{figure} 
\subsection{{Simulation 2: An inference example involving a structured MAR mechanism}}
{We now demonstrate how SM can affect inferences}, especially when dealing with strong structures. We begin by generating data for $n=1000$ subjects, {as follows}: 
\begin{align}
 X_1 &\sim N(0, 1) \nonumber \\
    X_2 \mid (X_1=x_1) & \sim N(2x_1, 1) \nonumber \\ 
      X_3 \mid (X_2=x_1) & \sim N(1+x_1+2x_2, 1). \label{pop} 
        \intertext{
We let $M_1$, $M_2,$ and $M_3$ denote the corresponding missingness indicator vectors for $X_1$, $X_2$, and $X_3$. We impose a {MAR} missing data mechanism on $X_2$ by supposing that values in $X_2$ depend on $X_1$,}
   p(M_{2}=1 \mid x_1 )&=\frac{\text{exp}(2x_1)}{1+\text{exp}(2x_1)}.  \label{mechsimMAR} 
     \intertext{We then impose a SM mechanism on $X_3$ by assuming that $M_3$ depends only on $M_2$:}
   p(M_{3}=1 \mid M_2)&=
\begin{cases}
  q & \text{if $M_{2}=0$ }  \\
     0 & \text{if $M_{2}=1$ }
  \end{cases}.
  \label{mechsim5} 
  \end{align}
  The DG in Figure \ref{sim1mechanism} illustrates the relationships between $X_1$, $X_2$, $X_3$,$M_1$, $M_2$, and $M_3$. \par
    \begin{figure}[h!]
  \centering
        \includegraphics[width=7cm]{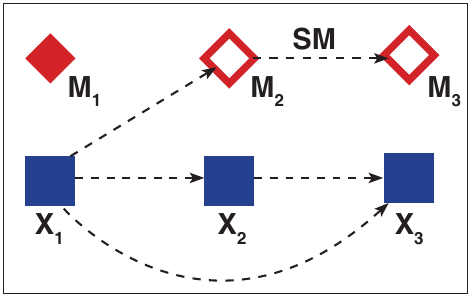}
\caption{\label{sim1mechanism} The causal effects present in Simulation 2. } 
\end{figure}
  By considering a range of $q$ (from 0 to 1) for each simulated data set, we can assess the effect of structure on inferential validity. The parameter $q$ links to the notion of weak and strong SM structure. When $q=0$, there is no effect of $M_2$ on $M_3$, representing an unstructured mechanism (MAR-U). When $q \in (0,1)$, the effect of $M_2$ on $M_3$ represents weak structure (MAR-WS). And when $q=1$, the effect of $M_2$ on $M_3$ represents strong structure (MAR-SS). \par 
{We again multiply impute missing values using 
\pkg{mice}.} Apart from switching to Bayesian normal linear regression imputation models (\texttt{method=`norm'}) to accurately reflect the true relationship in (\ref{pop}), we use \pkg{mice}'s default settings, which includes the generation of $m=5$ imputed data sets (\texttt{m=5}). The default setting for the algorithm's number of iterations is \texttt{maxit=5}. To show that this is insufficient for larger values of $q$, we also repeat the simulation with \texttt{maxit=50}.  \par 
     We consider the following analysis model, a normal linear regression model of $X_2$ on $X_1$:
\begin{align}
    X_{3}&=\beta_0+ \beta_1 X_1 + \beta_2 X_2 + \varepsilon, \nonumber
\end{align}
     where $\varepsilon$ is a $N(0, \sigma^2)$ random variate. We know from the formulation in (\ref{pop}) that the true, underlying values for these regression coefficients are $\boldsymbol\beta = (\beta_0,\beta_1, \beta_2) = (0,1,2)$. After fitting this model to each imputed data set and applying the multiple imputation (MI) combining rules \citep{Rubin1987}, we compute the {bias} to assess the validity of estimates, and the coverage to assess the proportion of 95\% confidence intervals that cover the true value.  \par The left and right plots of Figure \ref{sim1} give the bias and coverage when estimating the regression coefficient $\beta_2$. As $q$ tends towards 1, that is, as the structure becomes stronger, bias is introduced and the coverage proportion consequently falls away. When \texttt{maxit=5} (red circles), the bias increases at a faster rate than when \texttt{maxit=50} (turquoise triangles), showing that in this particular example the bias can partly be attributed to slow convergence. For example, when $q=0.9$, there is noticeable bias and undercoverage when \texttt{maxit=5}, but not for when \texttt{maxit=50}. When $q=1$, however, the bias cannot be reduced by increasing the number of iterations: there is a fundamental obstacle here that cannot be overcome. When $q = 1$, the strong structure means we have a file matching pattern, that is, $X_2$ and $X_3$ are never simultaneously observed, so there is no information on the true relationship between $X_2$ and $X_3$ (see \citealp{VanBuuren2018}). Hence the \pkg{mice} algorithm will never converge to the true value, even if we were to greatly increase the number of iterations. In general, as $q$ approaches 1 (but, importantly less than 1) and the fraction of missing information increases, a larger number of iterations is required – considerably more than the default of \texttt{maxit=5} – for the algorithm to converge.  \par The maximum percentage of missing information in $X_3$, which is approximately 50\%, occurs when $q=1$. To show that the bias present when $q=1$ is due to SM rather than an increasing of missingness percentage, we also ran the simulation where missingness in $X_3$ is determined through the same MAR mechanism used to impose missingness in $X_2$ (equation \ref{mechsimMAR}). We found that, in this instance, after values were multiply imputed using \pkg{mice} with \texttt{maxit=50}, estimates were still unbiased and confidence intervals valid. Thus, we can be sure that when $q=1$ the bias can be attributed to SM.
     \par This example highlights two statistical issues to be careful of when dealing with SM: an inherent, occasional inability to obtain valid inferences with strong structures, and slow convergence. With regards to the former, in this small scale example it is fairly clear how the bias arises – a lack of information. In large complex data sets, however, especially when undertaking multivariate analyses, this source of bias may not be obvious. Similarly, in this small scale example, it is straightforward to increase the number of iterations to allow the algorithm to converge. Yet in larger data sets this could be problematic. Firstly, assessing convergence of imputations and inferences is non-trivial in a multivariate space; and secondly, achieving convergence with SM may not scale well to high-dimensional data, requiring algorithms to be run for more iterations. Thus, in practice, when faced with SM, assessing convergence of either \pkg{mice} or other computational methods will likely involve a non-trivial decision around the number of iterations to run the method for, as well as careful inspection of convergence diagnostics.
  \begin{figure}[h!]
\centering
        \includegraphics[width=7.2cm]{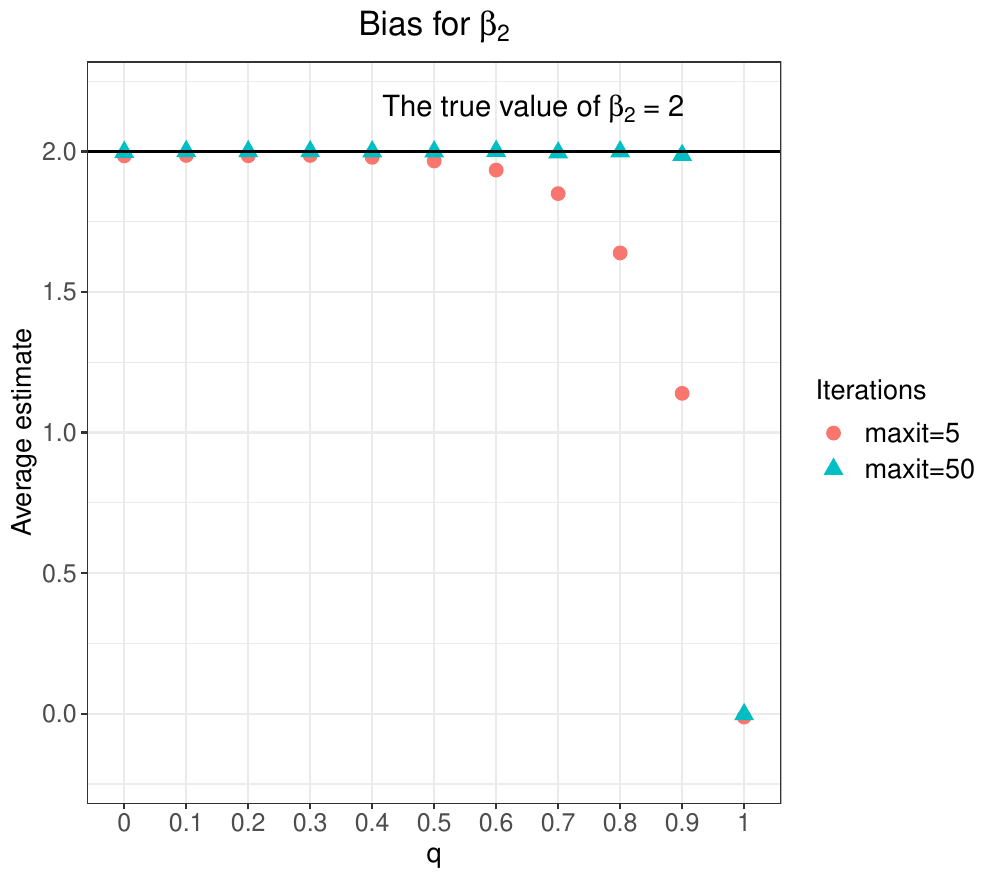}
        \includegraphics[width=7.2cm]{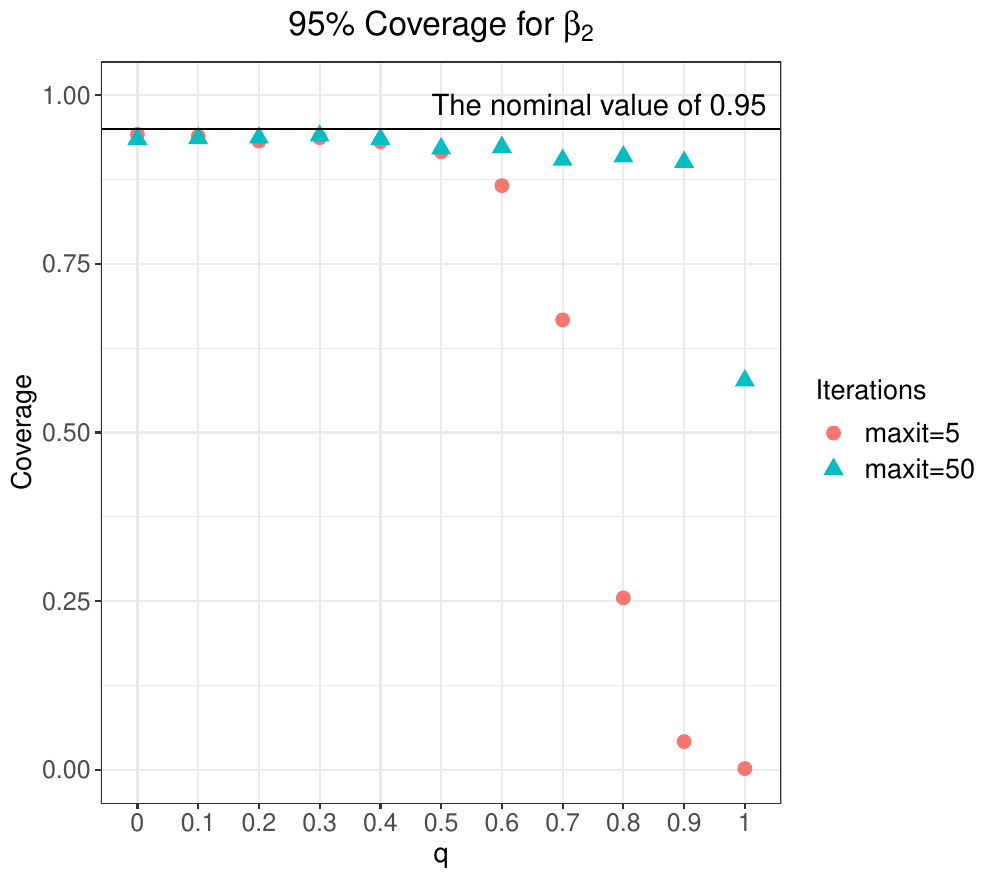}
        \caption{\label{sim1} The bias and coverage for $\beta_2$ for various $q$ and for 5 and 50 iterations.} 
\end{figure} 
\subsection{Simulation 3: A structured MNAR example}
{In this final simulation, we show how SM can be utilised to our advantage. Specifically, we consider a structured MNAR setting, and generate $n_\text{sim}=1000$ simulated data sets (with $n=1000$) as follows:} 
\begin{align}
 Z &\sim N(0, 1) \nonumber \\
    X_1 \mid (Z=z) & \sim N(2z, 1) \nonumber \\
        X_2 \mid (Z=z, X_1=x_1) & \sim N(1+z+2{x_1}, 1), \label{pop2} 
        \intertext{
and we let $M_1$ and $M_2$ denote the corresponding missingness indicator vectors for $X_1$ and $X_2$. We now suppose that we have an unobserved variable $Z$, that is, a latent variable that the analyst (imputer) does not have access to. We suppose that $Z$ has an effect on missingness in $X_1$, thus imposing an MNAR missing data mechanism on $X_1$,}
   p(M_{1}=1 \mid Z )&=\frac{\text{exp}(2z)}{1+\text{exp}(2z)}.  \label{mechsim1} 
     \intertext{We then impose a SM mechanism on $X_2$, by assuming that $M_2$ depends only on the previously observed $M_1$:}
   p(M_{2}=1 \mid M_1)&=
\begin{cases}
  1/2 & \text{if $M_{1}=1$ }  \\
     q & \text{if $M_{1}=0$ }
  \end{cases}
  \label{mechsim2} 
  \end{align}
  Essentially, the difference between the setup here and that used in the previous simulation is that, whereas previously $Z$ was observed, now it is missing. The DG in Figure \ref{sim2mechanism} illustrates the setup for this example, which is nearly identical to that in Figure \ref{sim1mechanism}. \par
  The parameter $q$ again links to the notion of weak and strong SM structure: when $q=0$, we have an MNAR-U mechanism; when $q \in (0,1)$, we have an MNAR-WS mechanism; and when $q=1$, we have an MNAR-SS mechanism. Moreover, in this example when $q \in (0,1/2)$, there is a positive SM mechanism between $M_1$ and $M_2$, that is, missingness in $M_1$ increases the probability of missingness in $M_2$; and when $q \in (1/2,1)$, there is a negative SM mechanism between $M_1$ and $M_2$, that is, missingness in $M_1$ reduces the probability of missingness in $M_2$. \par
   \begin{figure}[h!]
  \centering
        \includegraphics[width=7.2cm]{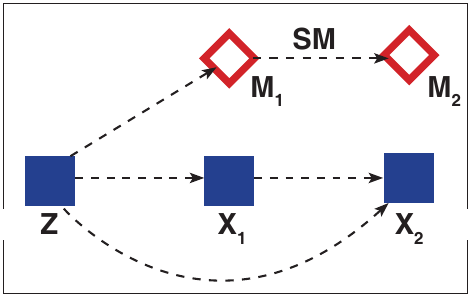}
\caption{\label{sim2mechanism} The causal effects present in Simulation 3. } 
\end{figure}
The estimand we consider is the expectation of $X_2$, for which the true value is 0. As mentioned earlier, we suppose that the imputer does not have access to the latent variable $Z$. We consider two methods of imputation for dealing with missing values:
\begin{enumerate}
    \item 
    \begin{align*}
        \text{Impute}\; X_1 \; \text{conditional on} \; X_2 \\
        \text{Impute}\; X_2 \; \text{conditional on} \; X_1
    \end{align*}
    This first method requires the use of FCS. We run 50 iterations (\texttt{maxit=50}) of the \pkg{mice} algorithm, with \texttt{method=`norm'} and \texttt{m=5}. This can be viewed as the standard way of multiply imputing missing values.
    \item 
    \begin{align*}
        \text{Impute}\; X_2 \; \text{conditional on} \; M_1 
    \end{align*}
    As the estimand of interest – the expectation of $X_2$ – does not depend on $X_1$, we do not need to impute missing values for $X_1$. We can instead use the SM relationship between $M_1$ and $M_2$ to impute missing values for $X_2$ using a model that depends on $M_1$ (not $X_1$). A secondary benefit of this approach is that, by definition, $M_1$ only includes 0s and 1s, so {is completely observed} and thus FCS is not required to impute missing values for $X_2$.
    \end{enumerate} \par
The results are given in Figure \ref{simMNAR}. Approach (a), denoted by the red circles, {clearly fails, with estimates hovering around 0 instead of 1 (the estimates do improve slightly as $q$ increases, and association between $M_1$ and $M_2$ helps to more accurately capture the relationship between $X_1$ and $X_2$). Owing to this bias, the confidence intervals can never cover the true value. The results for approach (b), on the other hand, denoted by the turquoise triangles, are clearly unbiased and the coverage values are at the nominal level.} \par Thus, in this example, if we replace $X_1$ with the missingness indicator $M_1$ in the imputation model for $X_2$, we obtain valid inferences. In this instance, we are utilising the fact that missingness in $X_2$ depends on \textit{missingness} in $X_1$ – that is, $M_1$ rather than $X_1$ – and by removing the link between $X_1$ and $X_2$ we are avoiding bias caused when imputing $X_1$ from propagating through to $X_2$. {While this is a relatively simple example, it demonstrates the advantages possible from leveraging information/structure present in $\mathbf{M}$ – in addition to the information in $\mathbf{X}$ – to best address problems posed by the missing data.}    
  \begin{figure}[h!]
\centering
            \includegraphics[width=7.2cm]{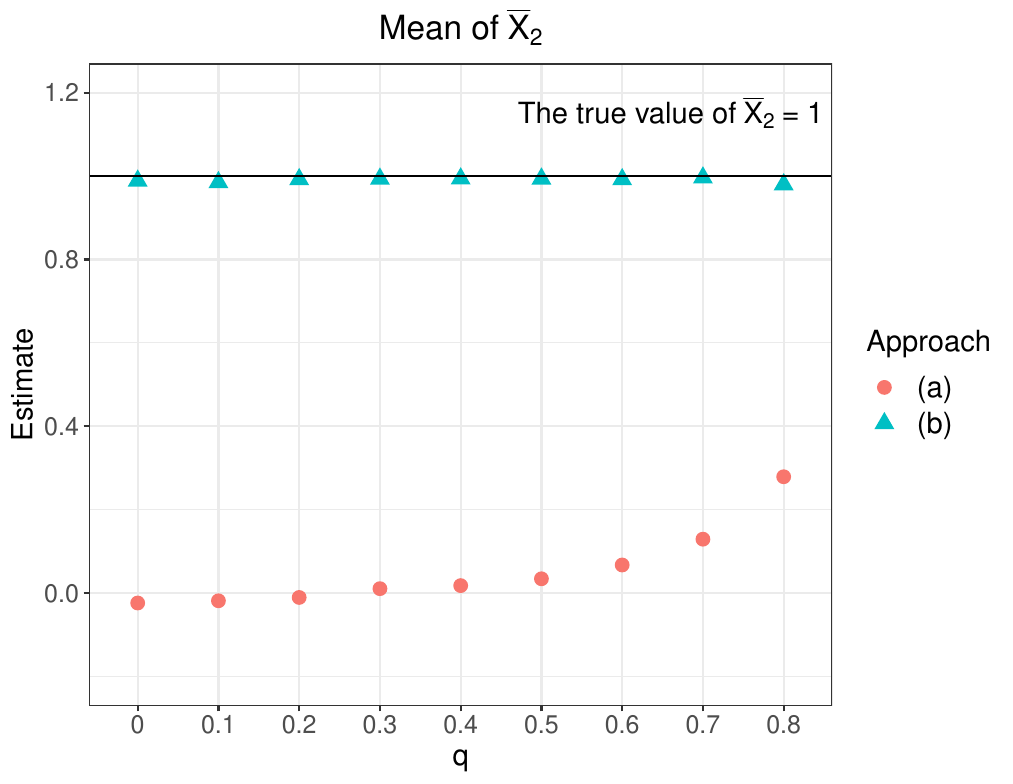}
            \includegraphics[width=7.2cm]{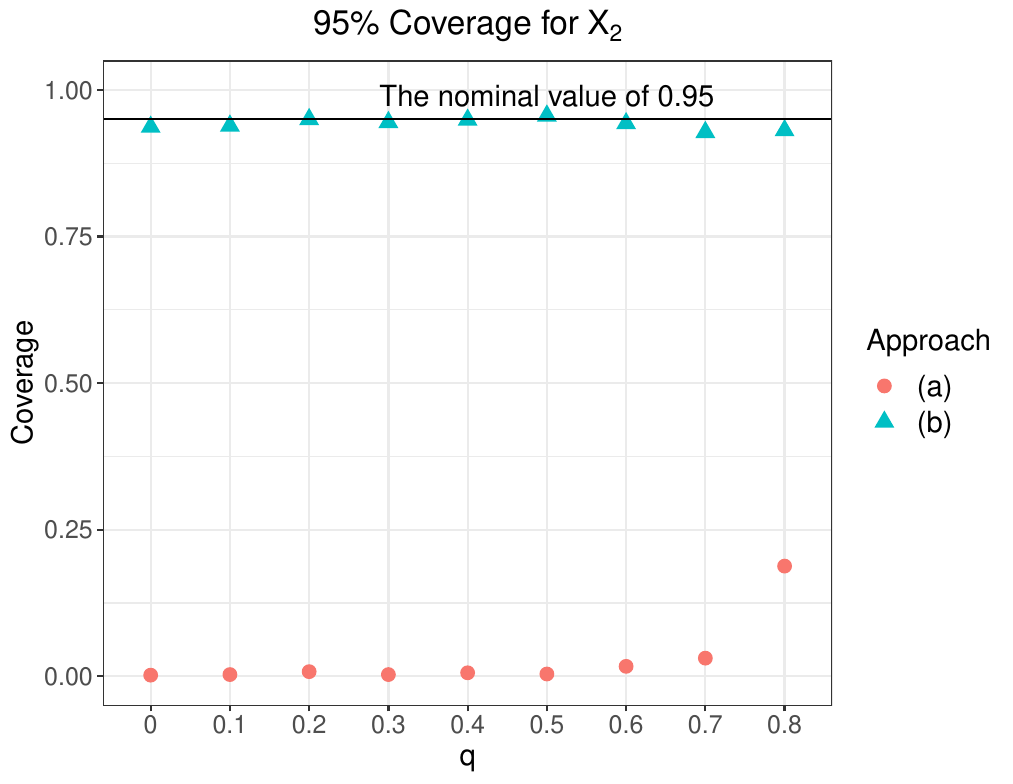}
        \caption{\label{simMNAR} The estimates (left) and coverages (right) for $\bar{X}_2$ in the third simulation.} 
\end{figure} 
\section{Structured Missingness within Real-World Clinico-Genomic Databases (CGDB)}  \label{sec9}
We now consider a real-world oncology clinico-genomic database (CGDB), formed through the linkage of Flatiron Health (FH) electronic health records (EHR) with Foundation Medicine (FMI) comprehensive genomic profiling for patients with cancer in the United States treated at approximately 280 US cancer clinics ($\sim$800 sites of care) \citep{Singal2017}. The database consists of 22 individual CGDBs, 21 of which are disease-specific and one which is disease-agnostic. In each, retrospective longitudinal clinical data are derived from FH EHR data, comprising patient-level structured and unstructured data, curated via technology-enabled abstraction, and linked to genomic data derived from FMI comprehensive genomic profiling tests by de-identified, deterministic matching \citep{Birnbaum2020}.

Together, the CGDBs represent an impressive collection of longitudinal, patient-level data encompassing over 100,000 patients diagnosed with cancer. This comprehensive data source provides scientists with an invaluable resource to study not only cancer-specific cohorts in each disease-specific CGDB, but also to leverage the collection of CGDBs across cancers to explore pan-tumor or tumor-agnostic insights, a recently emerging paradigm in cancer treatment based on shared molecular characteristics across cancer types \citep{Cristescu2022, Doebele2020, Becker2020}. This research advantage arises from the CGDB's ability to offer rich and diverse datasets, enabling the study of commonalities and differences across cancers at both clinical and genomic levels. By aggregating information from a vast number of patients, it is possible to explore shared molecular characteristics, treatment responses, and potential biomarkers that transcend individual cancer types. Such a `pan-tumor' approach can uncover novel therapeutic strategies and guide personalized medicine approaches, transforming cancer research and ultimately improving patient outcomes.

Collating these data across patients with different cancer types, each with different sets of clinically-relevant measurements, can give rise to SM challenges purely as a consequence of data combining. Furthermore, SM may be inherent in data collection and batch testing, for example, panels of lab tests or genomic tests. Such instances of SM can pose analytical challenges when seeking to learn from the totality of the CGDB, and should be characterised. We discuss several motivating examples in the sub-sections that follow.  

Table \ref{CGDBsummary} gives a brief summary of the variables {used to illustrate SM} in this section. \par 
\begin{table}
\caption{\label{CGDBsummary} A summary of variables derived from a tumor-agnostic cohort leveraging multiple disease-specific CGDBs.}
\centering
\begin{tabular}{*{2}{l}}
Name & Description  \\ \hline
CANCER TYPE & Patients' cancer types {(e.g., breast, prostate, ovarian)} \\ 
YEAR OF BIRTH & Patients' year of birth, used to derive ages \\
SEX & Patients' biological sex \\
{DATE} & {Tumor specimen collection date} \\
GENOMIC TEST & The type of genomic test administered \\
BAIT SET & The set of genes {tested (auxiliary variable)} \\
BRCA1/2, BCL10 and CASP8 & {Different genes} \\
PSA$_{1}$, PSA$_{2}$, PSA$_{3}$ & Series of PSA test results \\
\hline 
\end{tabular}
\end{table} 
\subsection{Example 1: PSA Testing }
The prostate-specific antigen (PSA) is a {common blood} biomarker {for the diagnosis, screening and monitoring of} prostate cancer{. Levels of PSA tend to be abnormally elevated in patients with prostate cancer and are measured because of their association with prostate cancer severity. PSA tests are therefore routinely administered to male patients with prostate cancer every 6–12 months, in particular following radiation therapy or surgery. Undiagnosed males who are high-risk (for example, carrying a mutation to the BRCA2/BRCA1 genes) may also receive PSA screening regularly. However, females, lacking a prostate, do not receive this test.} Therefore, for sensible reasons, {the PSA test is absent in the records of female patients and it is commonly missing among} male patients {diagnosed} with another type of cancer. 

For the sake of illustration, let us evaluate this scenario in the context of the SM taxonomy introduced in Section \ref{sec4}. Suppose we wish to evaluate a time series of PSA tests across lab visits for patients in a pan-tumor research setting where the disease-specific CGDBs have been combined across cancer types and patients, including males and females. Firstly, in the case of female patients, sex has a deterministic effect on whether {a series of PSA tests} are missing. This represents a {MAR-UD} mechanism. It can also be considered a case of logical missingness, as clearly a PSA test would be uninformative and inappropriate in this instance. The left plot of Figure \ref{PSA} shows this scenario graphically; SEX has a deterministic effect (solid arrows) on the missingness indicators (red triangles) for the PSA variables, PSA$_{1}$, PSA$_{2}$, and PSA$_{3}$. 

Secondly, we consider male patients with cancer types other than prostate cancer; these patients may receive PSA tests {as part of routine screening} based on risk factors. Those with certain risk factors – such as older patients or those with mutations to the BRCA2/BRCA1 genes – are more likely to be tested. In particular, the BRCA2 mutation has more recently been identified as a strong risk factor for prostate cancer diagnosis and severity, leading to calls by the research community to screen PSA levels earlier in affected men \citep{Page2019}. If these risk factors are observed, they would have a probabilistic effect on whether a patient receives a PSA test on a given visit, hence the dashed causal arrows from YEAR OF BIRTH and BRCA1/2 to the missingness indicators for PSA$_{1}$, PSA$_{2}$, and PSA$_{3}$ in the right plot of Figure \ref{PSA}. 
There is also an element of structure in the missingness, too, because if {the test} returns {normal values for a patient}, there is less need to repeat the test on the next visit, especially if little time has elapsed between the visits. In some instances, therefore, there is a negative SM relationship between the PSA missingness indicators (denoted by arrows between the PSA missingness indicators in the right plot of Figure \ref{PSA}), as observing the test at one visit increases the probability of missing it at the next visit (and \textit{vice versa}). This mechanism can then be viewed as a sequential MAR-WS. \par
 \begin{figure}[h!]
  \centering
                \includegraphics[width=7.2cm]{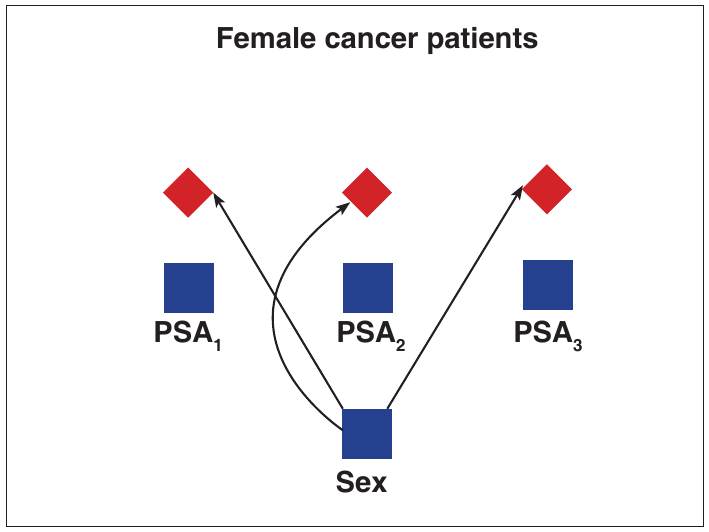}
                  \includegraphics[width=7.2cm]{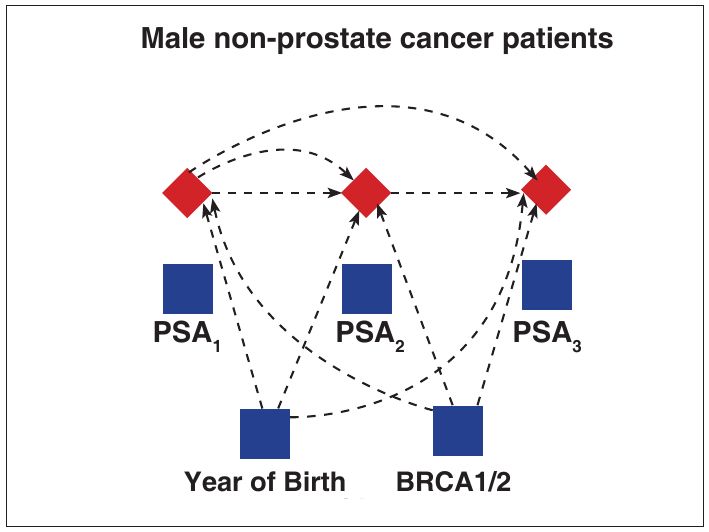}
\caption{\label{PSA} DGs relating to the PSA testing example;{ see Table \ref{CGDBsummary} for a summary of the variables. The blue squares denote the variables' values (elements of the matrix $\mathbf{X}$) and the red triangles denote the variables' missingness indicators (elements of the matrix $\mathbf{M}$).}} 
\end{figure}
\subsection{Example 2: Genomic Testing }
{Before considering this next example, it is worth noting the unit non-response within the CGDB in relation to the population of cancer patients more widely. The CGDB only includes cancer patients {in the FH EHR database} who {received comprehensive} genomic profiling {testing to characterise the DNA or RNA alterations that may be driving the growth of a specific tumor}. Hence the individuals included represent a sample of a larger population of patients with cancer. {The decision to test is} made by the physician, dependent on many factors, including: the current disease state of the patient, {family history,} their response to previous therapies, and any {disease presentation} indicating the likelihood of their cancer being driven by a particular gene mutation. It can be argued, therefore, that these missing records – relating to patients with cancer who did not undergo genomic profiling, {thus not observed in the CGDB} – are MAR-UP. This would need to be accounted for if obtaining inferences about the population as a whole.} 

{Patients included in the CGDB have received {one or more comprehensive genomic profiling tests, each measuring some set of cancer-relevant} genes (``bait sets'') {ranging from dozens to hundreds of genes}. Genomic alterations are identified via comprehensive genomic profiling (CGP) of cancer-related genes on FMI's next-generation sequencing (NGS) tests \citep{Woodhouse2020, He2016, Frampton2013}. {The choice of genomic test} is {influenced by multiple} factors, {including} the type of cancer {(for example, solid vs. haematologic malignancies).} Furthermore, the date of the test will dictate the particular bait set {(that is, the list of genes) used as an assay which may change over time (so a particular genomic test may assay different bait sets depending on when the test is performed). For example, the gene BCL10, commonly mutated in B-cell lymphomas, is tested exclusively in the haematologic bait sets. Conversely, the gene CASP8, which is known to be mutated in a number of solid tumors, is tested exclusively in the solid tumor bait sets. 

These multigene panels include guideline-recommended genes relevant for oncology, and a typical analysis could use a disease-specific CGDB to study the relevant disease cohort of interest. However, generating insights across cancer types (that is, solid and haematologic malignancies) will require the combining of multiple cancer cohorts, with different bait sets used in each cohort, that can exacerbate SM in the genomic data. Thus these broader, pooled cohorts may exhibit SM as a result of test (and potential changes in standard of care) dictating the measurement of genes, rather than biology. 

For example, in the case of analysing genomic alterations in haematological and solid tumor malignancies together, patients with diffuse large B-cell lymphoma (haematological) might be systematically missing the alteration status of CASP8 and patients with breast cancer (solid tumor) might be systematically missing BCL10. Looking exclusively at the genomic data, the missingness of these genes may appear to be a function of CANCER TYPE and DATE: in other words,} the missingness mechanism for the gene variables {BCL10 and CASP8} can be viewed as MAR-UP, where CANCER TYPE and DATE have a probabilistic effect on whether these genes are missing.} 

\par Alternatively, however, an auxiliary variable can be introduced here, say BAIT SET or GENOMIC TEST, that denotes the type of bait set and genomic test a patient received, respectively. These variables would otherwise be considered nuisance variables; they would not, for example, be included in a model for prediction. Nevertheless, they would introduce strong structure into the missingness mechanisms for the gene variables because SM can occur at either the test level or the bait set level; for example, BCL10 exhibits SM at the GENOMIC TEST level because it is only included in bait sets for certain haematologic tests and not others. On the other hand, CASP8 exhibits SM at the BAIT SET level. Both occurrences of SM may be influenced by CANCER TYPE and DATE, affecting the genomic test and bait set administered, and then, if a given test or bait set is missing, the genes measured by them are also missing. {Two examples are given in the DGs in Figures \ref{DGexamples1} and \ref{DGexamples2}. In Figure \ref{DGexamples1}, which excludes the bait set or genomic test variable, CANCER TYPE and DATE are shown to have a direct effect on the missingness indicators (red triangles) for the gene of interest, BCL10 and CASP8. In Figure \ref{DGexamples2}, CANCER TYPE and DATE affect the missingness indicators for the BAIT SET or GENOMIC TEST variables, which in turn have a deterministic effect on missingness in BCL10 and CASP8, respectively.}  

\par This example, where introducing an auxiliary variable changes the dynamic of the nature of the missingness mechanism, poses an interesting research question: which representation is to be preferred? While, on one hand, omitting the auxiliary variable makes the missingness mechanism easier to model – for example, omitting BAIT SET means that the mechanism is MAR in the sense of \cite{Rubin1976}, so multiple imputation can be utilised – on the other hand, including the auxiliary variable, which can be considered a mediator variable, allows a deeper understanding of the true underlying process at work. {The answer to the above question, therefore, likely depends on the analysis being performed, and the analyst's motivation for trying to understand the underlying structure of the missingness.} 
 \begin{figure}[h!]
  \centering
        \includegraphics[width=7.2cm]{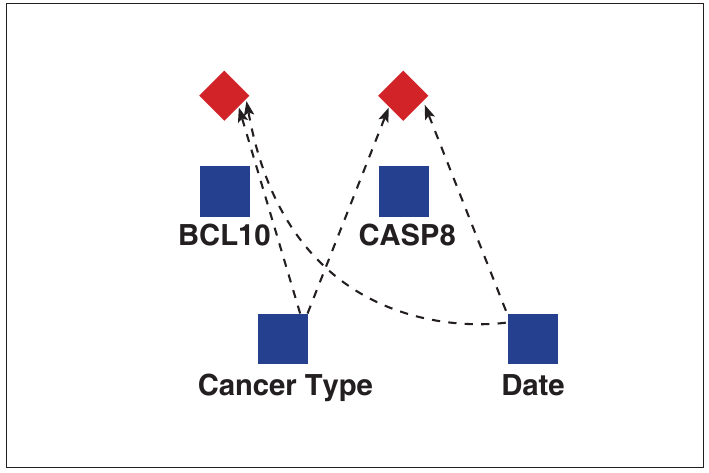}
\caption{\label{DGexamples1} Viewing the genomic testing example without considering SM. } 
\end{figure}
\begin{figure}[h!]
  \centering
        \includegraphics[width=7.2cm]{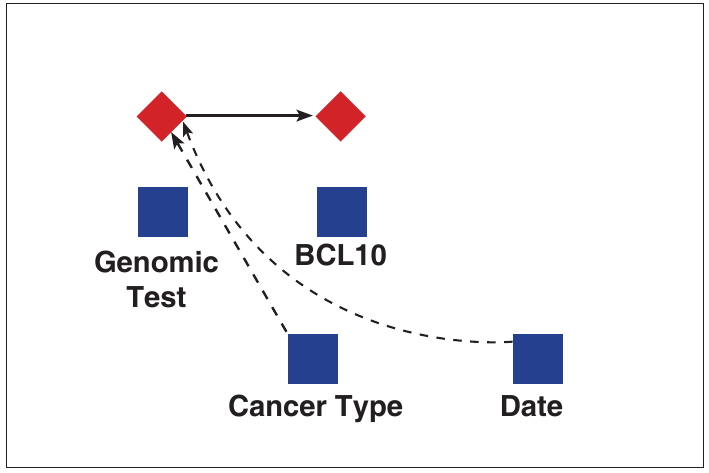}
                \includegraphics[width=7.2cm]{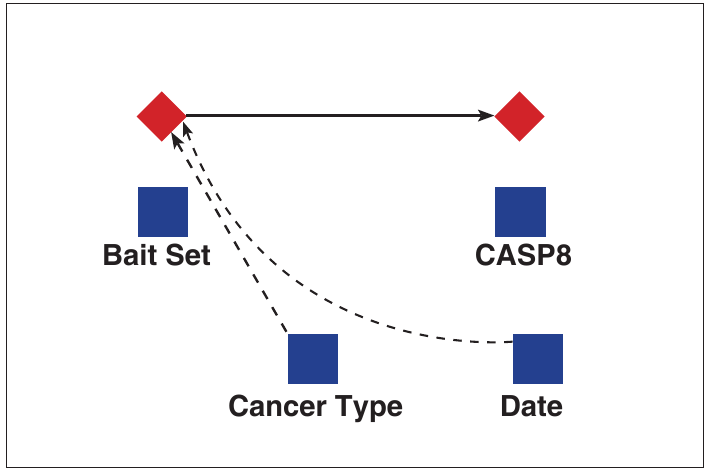}
\caption{\label{DGexamples2}  Viewing the genomic testing example in terms of SM. In the left example, missingness in GENOMIC TEST implies missingness in BCL10 with certainty. In the right example, missingness in BAIT SET implies missingness in CASP8 with certainty. } 
\end{figure}
\section{Discussion} \label{sec10}
Given that SM encompasses such a wide range of missingness scenarios, setting out a new taxonomy provides a solid foundation for characterising it, and thereby allowing appropriate action {to be taken}. {Exploration of the missingness indicator matrix $\mathbf{M}$ -- whether it is its effect on inferences' validity or as a source of information from which to improve the quality of inferences -- has, for no obvious reason, largely been overlooked in the literature. The matrices $\mathbf{X}$ and $\mathbf{M}$, of course, go hand-in-hand: $\mathbf{M}$ can be viewed as $\mathbf{X}$ maximally flattened to 0s and 1s. {The question which is fundamental to our paper is whether $\mathbf{M}$ is just a less informative representation of $\mathbf{X}$, or whether the multivariate relationships within $\mathbf{M}$ can help us to better understand the underlying mechanisms governing $\mathbf{X}$?} \par {This paper begins to address the {nine} Grand Challenges set out by \cite{Mitra2023} in relation to SM. In addition to defining SM (Challenge 1), we begin, through the use of DGs, to also consider the relationship between SM and causality (Challenge 8). Arguably more importantly, however, is the fact that by defining SM we are helping to facilitate further research into SM. This taxonomy will assist, for example, in understanding the geometry of SM (Challenge 2), as we may expect strong (deterministic) structures to exhibit sharply defined block missingness patterns. Moreover, this taxonomy should allow us to design experiments to mitigate any deleterious impact of SM (Challenge 3), will clearly help in devising methodological approaches for prediction or inference (Challenges 4--6), and will provide a starting point for developing benchmark data sets for {SM} (Challenge 7). Finally, understanding the underlying mechanisms for SM, and their impact on the observed data and any inferences made from that data, will inform on the risk of any sociocultural biases (Challenge 9).} 
\par As mentioned in Section \ref{com}, in practice, multiple mechanisms are likely to be present in any one data set, especially in complex {multi-modal} data sets such as the CGDB. Further research could be focused on considering how such mechanisms interact. If there is weak structure (a probabilistic relationship) between $\mathbf{M}$ and $\mathbf{X}$, for example, but a strong structure (a deterministic relationship) between ${M}_{j}$ and $\mathbf{M}_{-j}$, then how would these two structures interact? Would the strong structure ``dominate'' the weak structure? That is, would the probabilistic relationship be irrelevant given the presence of a deterministic relationship?

{{}{

{As we seek to develop methods that address and utilise SM in practice, it is important we keep in mind that the missing data themselves are not typically of primary interest; rather, it is the analysis of the complete data that drives key applied research questions, and the missing values are seen as a nuisance. As a result, we can postulate {the key questions applied researchers would want to address when faced with the potential of SM obstructing their analysis.}
\begin{enumerate}[{1)}]
\item Determine the (likely unknown) types of SM present in their data.
\item Identify which types of SM (amongst those identified in the data) are problematic for their desired analysis.
\item Implement relevant methods to deal with these types of SM to obtain valid analyses and extract maximum information from the data. 
\end{enumerate}
As mentioned earlier, certain approaches have considered aspects of SM in passing, and may be useful to consider developing further when seeking to address SM, in light of the above points.

{For 1), the suggestion of \cite{tierney2015} to utilise tree models to learn about structures present among the missing data is particularly appealing, especially when multiple (unknown) types of SM are present in the data. A range of approaches could be considered, from standard Classification and Regression Trees \citep{Breiman1984}, to more sophisticated approaches that incorporate uncertainty into the tree structure, such as Random Forests \citep{Breiman2001} or Bayesian Additive Regression Trees \citep{Chipman2010}.} 

{For 2), the concept of m-graphs given in \cite{mohan2021} could potentially be explored and developed into building causal relationships between the missing data indicators themselves within a framework that provides valid inferences for certain SM scenarios. 
More generally, the extensive literature on causal inference and graphical models will likely also prove useful to consider, particularly when SM can be characterised using purely causal pathways.} {Morever, while we have typically presented SM as an obstacle to overcome, in some instances the structure can be viewed as a help rather than a hindrance, as seen in Simulation 3 in Section \ref{sec8}. Thus, a complementary goal researchers may also like to consider here, is determining which types of SM can be leveraged to provide important information for feeding into their analysis.} 

{Lastly for 3), in large multivariate settings involving a number of unknown SM mechanisms, straightforward application of off-the-shelf methods for incomplete data, such as \pkg{mice}, or other standard multiple imputation packages, are not immediately obvious. {Shared traits between strong structure and block missingness suggest exploring the potential of existing methods proposed to impute blocks of missing data. For example, \cite{Li2014} propose an approach to multiply impute missing values through ordered monotone blocks.} More generally, Bayesian methods offer a great degree of flexibility, which could be utilised to develop a unifying framework; for example, modelling data with a mix of logical and random SM has been achieved through multi-level models \citep{audigier2018}, while \cite{Gelman1998} develop a Bayesian hierarchical regression model for data collated across several different surveys. Utilising the full Bayesian machinery to address SM is particularly appealing, whether this be model averaging inferences over multiple possible missing data mechanisms \citep{siddique2012} or leveraging information provided by SM through the use of informative priors.} 

To conclude, research into SM is critical, timely, and a necessary component to unlocking the full potential associated with large complex databases. We thus hope this contribution stimulates interest amongst the statistics community, as well as the scientific community more generally, to develop theory and methods that address the challenges, as well as the opportunities, posed by SM.}  
\bibliography{main.bib}
\end{document}